\documentclass[article,12pt]{elsarticle}

\biboptions{sort&compress}

\usepackage{amssymb}
\usepackage{amsmath}
\usepackage[version=4]{mhchem}
\usepackage{siunitx}
\usepackage{array}
\usepackage{array}  
\newcolumntype{L}[1]{>{\raggedright\arraybackslash}p{#1}}
\usepackage{tabularx}
\usepackage{longtable}
\usepackage{caption}
\usepackage{subcaption}
\usepackage{float}
\usepackage{overpic}
\usepackage{seqsplit}
\usepackage{tikz}
\usetikzlibrary{arrows.meta}
\usepackage{url}

\newcommand{\fld}[1]{\texttt{\small\expandafter\seqsplit\expandafter{\detokenize{#1}}}}

\journal{Scientific Data}

\begin{document}

\begin{frontmatter}

\title{A database of optical and thermal-radiative behaviour of materials under solar radiation}

\author[ntua]{Vasiliki E. Alexopoulou\cormark[1]}
\author[ntua]{Nikolaos Karalis}
\author[ntua]{Konstantinos Asimakopoulos}
\author[comoti]{Andrei Totu}
\author[comoti]{Constantin Sandu}
\author[ntua]{Paul-Peter Sotiriadis}

\affiliation[ntua]{organization={Department of Electrical and Computer Engineering, National Technical University of Athens},
            addressline={9, Iroon Polytechniou str.}, 
            postcode={15772},
            city={Zografou}, 
            state={Athens},
            country={Greece}}

\affiliation[comoti]{organization={Aerospace Propulsion/ Aeroacoustics/ Physics of Flight/ Gravity/ Environment, National Research and Development Institute for Gas Turbines -- COMOTI},
            addressline={220D, Iuliu Maniu Ave}, 
            postcode={061126},
            city={Bucharest 6}, 
            country={Romania}}

\cortext[1]{Corresponding author: valexopoulou@mail.ntua.gr (V. E. Alexopoulou)}

\begin{abstract}
Solar--matter interactions are important in many scientific and engineering applications. Modeling them needs consistent optical and thermal--radiative properties over the full solar spectrum. Here we present SpaderOptics, a database for 21 materials: metals, oxides, semiconductors, polymers, ceramics, and carbon. It compiles wavelength-dependent refractive indices and extinction coefficients from the literature. Each literature source is kept separate, together with its original conditions. One physics-based calculation chain turns these data into reflectivity, transmissivity, absorptivity, and emissivity for different incidence angles, slab thicknesses, and temperatures. Combined with the Air Mass Zero (AM0) solar spectrum, the properties give absorbed, reflected, and transmitted solar flux for concentrated sunlight. Equilibrium temperatures are given for both a theoretical lumped body and for a slab body, in which thermal conduction is considered. All data are limited to the solid state, up to a material-specific upper temperature limit. Currently, the database applies to extraterrestrial (AM0) sunlight only. A graphical App is also provided for exploring and visualizing the
database.
\end{abstract}

\end{frontmatter}

\section{Background and Summary}

Solar--matter interactions are important in solar-thermal energy conversion, photovoltaic and photonic technologies, high-temperature materials processing, thermal management, and radiation--material studies \cite{Kalogirou2009-lm}. Highly concentrated sunlight is also studied for future space systems, for example to heat and remove space debris. The European Union project ``Space debris removal system using concentrated Solar Rays (SPADER)'', in which we take part, develops such a system. This project motivated the database presented here.

Simulating these interactions needs consistent optical and radiative properties over the full solar wavelength range. These are the complex refractive index, $n+ik$, and the derived reflectivity, transmissivity, absorptivity, and emissivity. Several compilations exist. Palik's handbook \cite{Palik1998-jg} and the continuously maintained RefractiveIndex.INFO database \cite{Polyanskiy2024-cv} collect measured and calculated optical constants. Volumes 7--9 of the Thermophysical Properties of Matter (TPRC) Data Series \cite{Touloukian1970-qv,Touloukian1971-qv,Touloukian1972-qv} collect measured thermal-radiative properties. The EKHI database \cite{Gabirondo-Lopez2026-dw} makes these data open and FAIR (findable, accessible, interoperable, reusable).

Existing compilations provide extensive optical and thermal – radiative data, but these data originate from different literature sources and experimental conditions. Optical constants and radiative properties are often reported independently and over different spectral and experimental domains,
making their direct combination potentially inconsistent. Moreover, existing databases provide material properties only, but not ready-to-use quantities for broadband sunlight, such as absorbed, reflected, and transmitted solar flux. Getting these quantities needs extra calculations before the data can enter a model.

SpaderOptics addresses both limitations. It keeps one calculation chain from the reported $n(\lambda)$ and $k(\lambda)$ to solar fluxes and equilibrium temperatures. Every derived quantity is calculated from the same optical data with physics-based only relations and without introducing any material-specific fitted coefficients or empirical parameters. Solar fluxes are calculated for the extraterrestrial AM0 spectrum, so the database applies to sunlight outside the atmosphere (e.g. for space applications). Equilibrium temperatures are given for a lumped body and for a slab body, in which thermal conduction is considered. All data end at a material-specific upper temperature limit. Above this limit, the literature data and the models used in the database no longer describe the solid state. The provenance and spectral coverage of every literature source are kept.

\section{Methods}

\subsection{Overview of the calculation chain}
\label{sec:overview}

\begin{figure}[t]
\centering
\begin{tikzpicture}[
  box/.style={draw, rounded corners=2pt, align=center, font=\small, text width=4.6cm, minimum height=1.0cm, inner sep=4pt, execute at begin node={\hyphenpenalty=10000\relax\exhyphenpenalty=10000\relax}},
  wide/.style={box, text width=7.2cm},
  inp/.style={box, dashed},
  arr/.style={-{Stealth[length=2.2mm]}, thick}]
\node[wide] (nk) at (0,0) {Refractive index $n(\lambda)$ and extinction coefficient $k(\lambda)$ of each source};
\node[wide] (res) at (0,-1.7) {Cleaning and resampling on the AM0 wavelength grid};
\node[wide] (opt) at (0,-3.4) {Slab optics: $R$, $T_r$, $A$ for angle $\theta_i$, thickness $d$, and temperature $T$};
\node[box] (flux) at (-3.1,-5.6) {Solar flux $Q$: properties $\times$ AM0 spectrum, scaled by $C$ suns};
\node[box] (eps) at (3.1,-5.6) {Emissivity $\varepsilon$ (Kirchhoff's law) and emitted flux $Q_\mathrm{emit}$};
\node[box] (lump) at (-3.1,-8.0) {Lumped-body temperature from $Q_\mathrm{abs}=Q_\mathrm{emit}$};
\node[box] (slab) at (3.1,-8.0) {Slab temperatures $T_\mathrm{f}$, $T_\mathrm{b}$, $T_\mathrm{mid}$ with heat conduction};
\node[inp] (kap) at (3.1,-9.9) {Thermal conductivity $\kappa(T)$ from the literature};
\draw[arr] (nk) -- (res);
\draw[arr] (res) -- (opt);
\draw[arr] (opt.south) -- ++(0,-0.5) -| (flux.north);
\draw[arr] (opt.south) -- ++(0,-0.5) -| (eps.north);
\draw (flux.south) -- ++(0,-0.55) coordinate (a);
\draw (eps.south) -- ++(0,-0.55) coordinate (b);
\draw (a) -- (b);
\draw[arr] (a) -- (lump.north);
\draw[arr] (b) -- (slab.north);
\draw[arr] (kap) -- (slab);
\end{tikzpicture}
\caption{Calculation chain of the database. The chain is repeated for every literature source.}
\label{fig:chain}
\end{figure}
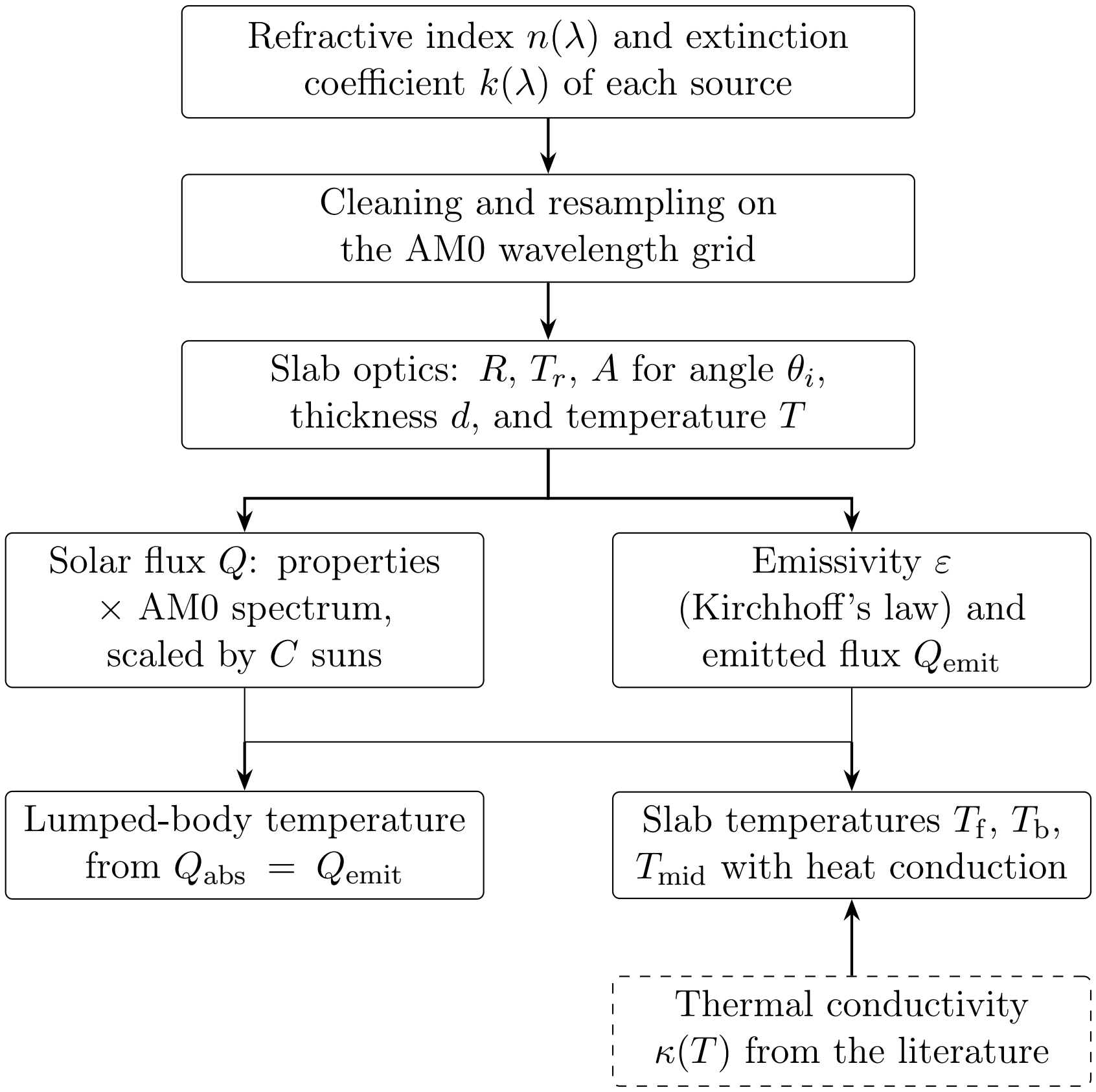

Figure~\ref{fig:chain} shows the calculation chain. It is repeated for every literature source.

\begin{enumerate}
\item The refractive indices and extinction coefficients, gathered from RefractiveIndex.INFO \cite{Polyanskiy2024-cv}, are cleaned and resampled on the AM0 wavelength grid (Sections~\ref{sec:nk}, \ref{sec:solar} and~\ref{sec:verification}).
\item The reflectivity, transmissivity, and absorptivity of a plane-parallel slab are calculated for different incidence angles, thicknesses, and temperatures (Section~\ref{sec:optics}).
\item The properties are weighted with the AM0 spectrum. This gives the absorbed, reflected, and transmitted solar flux, including concentrated sunlight (Section~\ref{sec:solar}).
\item Kirchhoff's law gives the emissivity and the emitted thermal flux (Section~\ref{sec:emissivity}).
\item Two equilibrium temperature models are used (Section~\ref{sec:temperature}). Lumped body model balances absorbed and emitted flux. Slab model also solves heat conduction through the thickness, using literature thermal conductivities.
\end{enumerate}

For each material, results end at a material specific upper temperature limit (Section~\ref{sec:upperlimit}).

Two terms are used throughout. ``Directional'' means evaluated for one incidence (or emission) angle $\theta_i$, measured from the surface normal. ``Hemispherical'' means averaged over all angles of one side of the surface, with the cosine-weighted average of Eq.~(\ref{eq:hemi}). We assume isotropic surfaces. Properties then depend on $\theta_i$ but not on the azimuth angle.

\subsection{Material selection}

The material selection was guided by the material groups of ECSS-Q-ST-70C Rev.~2 \cite{ECSS2019}, which are the main groups used in spacecraft. Debris fragments from collisions, explosions, ageing, or degradation come mostly from these groups. The 21 materials in Table~\ref{tab:materials} cover structural materials and their degradation products, such as oxides \cite{Fortescue2011,Wertz2001,Banks2003}. These classes are also widely used outside space applications.

\begin{table}[t]
\centering
\caption{Materials included in the database.}
\label{tab:materials}
\begin{tabular}{>{\raggedright\arraybackslash}p{0.33\textwidth}>{\raggedright\arraybackslash}p{0.6\textwidth}}
\hline
\textbf{Class} & \textbf{Materials} \\
\hline
Metals & aluminum (\ce{Al}), copper (\ce{Cu}), nickel (\ce{Ni}), titanium (\ce{Ti}), iron (\ce{Fe}), silver (\ce{Ag}), gold (\ce{Au}) \\
Oxides & alumina (\ce{Al2O3}), cuprous oxide (\ce{Cu2O}), ferric oxide (\ce{Fe2O3}), magnetite (\ce{Fe3O4}), titania (\ce{TiO2}) \\
Semiconductors and optical materials & silicon (\ce{Si}), silica (\ce{SiO2}), germanium (\ce{Ge}) \\
Polymers & Kapton-type polyimide, polymethyl methacrylate (PMMA) \\
Ceramics & silicon carbide (\ce{SiC}), titanium carbide (\ce{TiC}), zirconia (\ce{ZrO2}) \\
Carbon & amorphous carbon (\ce{C}) \\
\hline
\end{tabular}
\end{table}

Amorphous carbon represents carbon-fibre-reinforced polymers (CFRP) and carbon-based thermal protection systems. The optical response of a composite depends on the fibre orientation, the resin, and the manufacturing process \cite{Xu:15}. To keep the database at material level, we include carbon as a representative constituent. The polymer matrix is represented by the polymers in Table~\ref{tab:materials}.

\subsection{Refractive-index data}
\label{sec:nk}

The wavelength-dependent refractive index $n$ and extinction coefficient $k$ come from the YAML files of RefractiveIndex.INFO \cite{Polyanskiy2024-cv}. These files collect measured and theoretical data from the literature. The measurements use methods such as spectroscopic ellipsometry \cite{Cheng2016-cr}, reflectometry \cite{Larruquert1996-po}, and angle-averaged absorptance \cite{Ordal1988-me}. The theoretical data come from models such as the semi-quantum model \cite{Rakic1995-na}, the Lorentz--Drude model, and the Brendel--Bormann model \cite{Rakic1998-xo}.

Each material has one or more independent literature sources. We keep every source separate and do not merge them. In this way, the original spectral range and conditions stay visible, such as temperature, film thickness, and fabrication details. The reference and the url or doi of each source are also stored.

The optical data are stored as the complex refractive index,
\begin{equation}
\tilde{n}(\lambda)=n(\lambda)+ik(\lambda).
\label{eq:RefractiveIndex}
\end{equation}
Sources give tabulated $n$ and $k$, tabulated $n$ or $k$ alone, or analytical dispersion relations. When $n$ and $k$ are in separate blocks, we combine them on a common wavelength grid by linear interpolation. The analytical relations (nine types, listed in \cite{Polyanskiy2024-cv}) are evaluated on a uniform grid of 1000 points over the wavelength range of the source. 
If only $n(\lambda)$ is given, we set $k=0$. This is valid in transparent spectral regions, and the assumption is stored in the database.

Some sources give $n$ and $k$ at several temperatures. In RefractiveIndex.INFO, each temperature is a separate entry, and we keep these entries separate, too. For each such source, we also add one combined entry with the suffix ``-Tdepend'', which contains all temperatures in a combined array. To describe both cases (wavelegth-only dependence, and wavelength- and temperauture- dependence), we write
\begin{equation}
\tilde{n}(\lambda,T^\ast)=n(\lambda,T^\ast)+ik(\lambda,T^\ast),
\label{eq:Common_RefractiveIndex}
\end{equation}
where $T^\ast$ marks the temperature dependence when it exists. These complex refractive indices are the input to all calculations that follow.

\subsection{Optical properties of a slab}
\label{sec:optics}

\subsubsection{Notation and units}

Table~\ref{tab:symbols} lists the symbols and units. Subscripts ``dir'' and ``hemi'' mean directional and hemispherical (Section~\ref{sec:overview}). For example, $A_\mathrm{dir}$ is the absorptivity for one incidence angle (directional absorptivity), and $A_\mathrm{hemi}$ is the angle-averaged absorptivity (hemispherical absorptivity).

\begin{longtable}
	{>{\raggedright\arraybackslash}m{0.22\textwidth}
		>{\raggedright\arraybackslash}m{0.42\textwidth}
		>{\raggedright\arraybackslash}m{0.27\textwidth}}
	\caption{Symbols, quantities, and units.}
	\label{tab:symbols}\\
	\hline
	\textbf{Symbol} & \textbf{Quantity} & \textbf{Units} \\
	\hline
	\endfirsthead
	\hline
	\textbf{Symbol} & \textbf{Quantity} & \textbf{Units} \\
	\hline
	\endhead
	\hline
	\multicolumn{3}{r}{\textit{Continued on next page}}\\
	\endfoot
	\hline
	\endlastfoot
	$\lambda$, $\lambda_s$ & wavelength (on the AM0 grid for $\lambda_s$) & $\mu$m (m in the Planck's law) \\
	$\lambda_{s,\min}$, $\lambda_{s,\max}$ & limits of the common wavelength interval of the AM0 spectrum and the source & $\mu$m \\
	$\lambda^{\mathrm{AM0}}_{\min}$, $\lambda^{\mathrm{AM0}}_{\max}$ & limits of the AM0 wavelength grid & $\mu$m \\
	$n$, $k$ & refractive index, extinction coefficient & dimensionless \\
	$n_1$ & refractive index of the incident medium (vacuum, $n_1=1$) & dimensionless \\
	$n_j$, $k_j$, $n_\mathrm{new}$, $k_\mathrm{new}$ & $j$-th duplicate value and averaged value of $n$ and $k$ at one wavelength (Eq.~23) & dimensionless \\
	$N_i$ & number of duplicate entries at the wavelength $\lambda_i$ (Eq.~23) & -- \\
	$\tilde{n}=n+ik$ & complex refractive index & dimensionless \\
	$\theta_i$, $\theta_t$ & incidence (or emission) angle, transmitted angle, both from the surface normal & degrees (rad in calculations) \\
	$\varphi$ & azimuth angle & degrees (rad in calculations) \\
	$\mathrm{d}\Omega$ & element of solid angle, $\sin\theta_i\,\mathrm{d}\theta_i\,\mathrm{d}\varphi$ & sr \\
	$T$, $T^\ast$ & temperature & K \\
	$d$ & slab thickness & $\mu$m \\
	$L$ & lateral size of the slab ($L \gg d$) & $\mu$m \\
	$\omega$ & polarization ($s$ or $p$) & -- \\
	$r_s$, $r_p$ & Fresnel amplitude reflection coefficients & dimensionless \\
	$R_F$, $R_{F,s}$, $R_{F,p}$ & single-interface Fresnel reflectivity (unpolarized, $s$, $p$) & dimensionless (0--1) \\
	$R$ & reflectivity & dimensionless (0--1) \\
	$T_r$ & transmissivity & dimensionless (0--1) \\
	$A$ & absorptivity & dimensionless (0--1) \\
	$R_s$, $R_p$, $T_{r,s}$, $T_{r,p}$ & slab reflectivity and transmissivity for $s$ and $p$ polarization & dimensionless (0--1) \\
	$X$ & generic optical property ($R$, $T_r$, or $A$) in Eq.~(10) & dimensionless (0--1) \\
	$\tau$ & single-pass internal transmittance & dimensionless (0--1) \\
	$k_z$ & component of the wave vector along the thickness direction & $\mu$m$^{-1}$ \\
	$I_\mathrm{solar}$ & AM0 spectral irradiance & W m$^{-2}$ $\mu$m$^{-1}$ \\
	$q_{\mathrm{abs}}$, $q_{\mathrm{ref}}$, $q_{\mathrm{tr}}$ & spectral absorbed, reflected, transmitted irradiance & W m$^{-2}$ $\mu$m$^{-1}$ \\
	$Q_{\mathrm{abs}}$, $Q_{\mathrm{ref}}$, $Q_{\mathrm{tr}}$ & absorbed, reflected, transmitted solar flux & W m$^{-2}$ \\
	$Y$ & type of solar flux (abs, ref, or tr) in eqs.~(13) and (14) & -- \\
	$Q^{\mathrm{conc}}$ & solar flux scaled by the concentration $C$ (superscript conc) & W m$^{-2}$ \\
	$Q_{\mathrm{solar}}$, $Q_{\mathrm{AM0}}$ & AM0 irradiance inside the common wavelength interval, and in the complete AM0 spectrum & W m$^{-2}$ \\
	$C$ & solar concentration & dimensionless (suns) \\
	$\varepsilon$ & emissivity & dimensionless (0--1) \\
	$\varepsilon_{\lambda,\mathrm{dir}}$, $\varepsilon_{\mathrm{dir}}$, $\varepsilon_{\mathrm{hemi}}$ & spectral directional, total directional, and hemispherical emissivity & dimensionless (0--1) \\
	$B_{\lambda}$ & spectral radiance (blackbody, bb, or real body) & W m$^{-2}$ sr$^{-1}$ m$^{-1}$ \\
	$h$, $c$, $k_B$, $\sigma_B$ & Planck constant, speed of light, Boltzmann constant, Stefan--Boltzmann constant & J s, m s$^{-1}$, J K$^{-1}$, W m$^{-2}$ K$^{-4}$ \\
	$Q_{\mathrm{emit}}$ & emitted thermal flux & W m$^{-2}$ \\
	$\kappa$ & thermal conductivity & W m$^{-1}$ K$^{-1}$ \\
	$q_{\mathrm{cond}}$ & conductive heat flux through the slab & W m$^{-2}$ \\
	$T_\mathrm{eq}$ & lumped equilibrium temperature & K \\
	$\Delta T_\mathrm{eq}$ & change of the equilibrium temperature (ideal-mirror test) & K \\
	$\Delta Q_{\mathrm{dir\text{-}hemi}}$, $\Delta Q_{\mathrm{hemi\text{-}hemi}}$ & residual functions of the lumped-body energy balance & W m$^{-2}$ \\
	$T_\mathrm{f}$, $T_\mathrm{b}$, $T_\mathrm{mid}$ & front, rear, and mid-plane temperature of the slab & K \\
	$T_\mathrm{low}$, $T_\mathrm{up}$ & lowest and upper temperature of the temperature grid & K \\
	$\eta_s$, $\eta_{\mathrm{th}}$ & solar and thermal spectral coverage ratio & dimensionless (0--1) \\
	$\Delta x$ & absolute resampling error (of $n$ or $k$) & dimensionless \\
	$\delta_x$, $\delta_\mathrm{model}$ & relative resampling error, relative model error & \% \\
	$\lambda_0$ & limit wavelength of the thermal-emission share estimate (Eq.~27) & $\mu$m \\
	$f_{>\lambda_0}$ & share of the emitted power at wavelengths longer than $\lambda_0$ & dimensionless (0--1) \\
	$y$, $y_0$ & dimensionless variable $hc/(\lambda k_B T)$, and its value at $\lambda_0$ (Eq.~27) & dimensionless \\
	$E_b$ & emissive power of a blackbody & W m$^{-2}$ \\
	$tol_Q$ & energy-balance residual threshold & W m$^{-2}$ \\
	$\varepsilon_{\mathrm{calc}}$, $\varepsilon_{\mathrm{exp}}$ & calculated and measured emissivity & dimensionless (0--1) \\
	$N$ & number of sources of a material & -- \\
	$s$ & source index, $s=1,\dots,N$ (Section~4.5.4) & -- \\
	$\varepsilon_s$ & hemispherical emissivity of source $s$ & dimensionless (0--1) \\
	$\varepsilon_{\min}$, $\varepsilon_{\max}$ & minimum and maximum emissivity across the sources & dimensionless (0--1) \\
	$\bar{\varepsilon}$, $\varepsilon_{\mathrm{med}}$ & mean and median emissivity across the sources & dimensionless (0--1) \\
	$\sigma_{\varepsilon}$ & standard deviation of the emissivity across the sources & dimensionless \\
	$\mathrm{MAD}_{\varepsilon}$ & median absolute deviation of the emissivity & dimensionless \\
	$\Delta\varepsilon$ & absolute range of the emissivity, $\varepsilon_{\max}-\varepsilon_{\min}$ & dimensionless \\
	$\Delta\varepsilon_{\mathrm{rel}}$ & relative range of the emissivity, $100\,\Delta\varepsilon/\varepsilon_{\mathrm{med}}$ & \% \\
\end{longtable}

\subsubsection{Fresnel reflectance at a single interface}

We first calculate the response of one vacuum--material interface with the Fresnel equations \cite{BornWolf2019}. The wave comes from vacuum ($n_1=1$) and meets a semi-infinite material with complex index $\tilde n$. Snell's law \cite{Reider2016-ic} gives the transmitted angle $\theta_t$:
\begin{equation}
n_1\sin\theta_i=\tilde n\sin\theta_t,
\qquad
\cos\theta_t=\sqrt{1-\left(\frac{\sin\theta_i}{\tilde n}\right)^2}
\end{equation}
The Fresnel coefficients for the s and p polarizations are \cite{BornWolf2019}
\begin{equation}
r_s=\frac{\cos\theta_i-\tilde n\cos\theta_t}{\cos\theta_i+\tilde n\cos\theta_t},
\qquad
r_p=\frac{\tilde n\cos\theta_i-\cos\theta_t}{\tilde n\cos\theta_i+\cos\theta_t}
\end{equation}
The intensity reflectances are $R_{\mathrm{F},\omega}=|r_\omega|^2$ \cite{Reider2016-ic}, with $\omega=s,p$. Fresnel coefficients and intensity reflectances depend on $\lambda$, $T^\ast$, and $\theta_i$. For unpolarized light, the single-interface reflectivity is the average
\begin{equation}
R_{\mathrm{F}}=\tfrac{1}{2}\left(R_{\mathrm{F},s}+R_{\mathrm{F},p}\right)
\end{equation}

\subsubsection{Slab with incoherent internal reflections}

\begin{figure}[t]
\centering
\begin{tikzpicture}[>=Stealth, font=\small]
  \fill[gray!15] (0,0) rectangle (7,1.6);
  \draw (0,0) -- (7,0);
  \draw (0,1.6) -- (7,1.6);
  \node[anchor=west] at (0.15,0.8) {slab, $\tilde n$};
  \node[anchor=west] at (0.1,2.2) {vacuum};
  \node[anchor=west] at (0.1,-0.8) {vacuum};
  \draw[dashed] (2.8,-0.3) -- (2.8,3.1);
  \draw[->, thick] (1.0,3.1) -- (2.8,1.6);
  \node[anchor=south] at (1.0,3.1) {incident beam};
  \draw[->, thick] (2.8,1.6) -- (4.6,3.1);
  \node at (4.9,2.85) {$R$};
  \draw[->, thick] (2.8,1.6) -- (3.55,0);
  \draw[->, thick] (3.55,0) -- (5.3,-1.45);
  \node at (5.7,-1.15) {$T_r$};
  \draw[->, dashed] (3.55,0) -- (4.3,1.6);
  \draw[->, dashed] (4.3,1.6) -- (6.1,3.1);
  \node at (2.45,2.3) {$\theta_i$};
  \node at (3.45,0.8) {$\theta_t$};
  \node at (5.4,0.8) {absorption};
  \draw[<->] (6.7,0) -- (6.7,1.6) node[midway, right] {$d$};
  \draw[->, thick] (-0.6,0) -- (-0.6,1.9) node[above] {$z$};
\end{tikzpicture}
\caption{Sketch of the slab model. The $z$ axis is along the thickness $d$. A beam arrives at angle $\theta_i$ from the surface normal (dashed line). Part of it is reflected ($R$), part is absorbed inside the slab, and part is transmitted ($T_r$). Dashed arrows show one internal reflection. All internal reflections are summed.}
\label{fig:slab}
\end{figure}
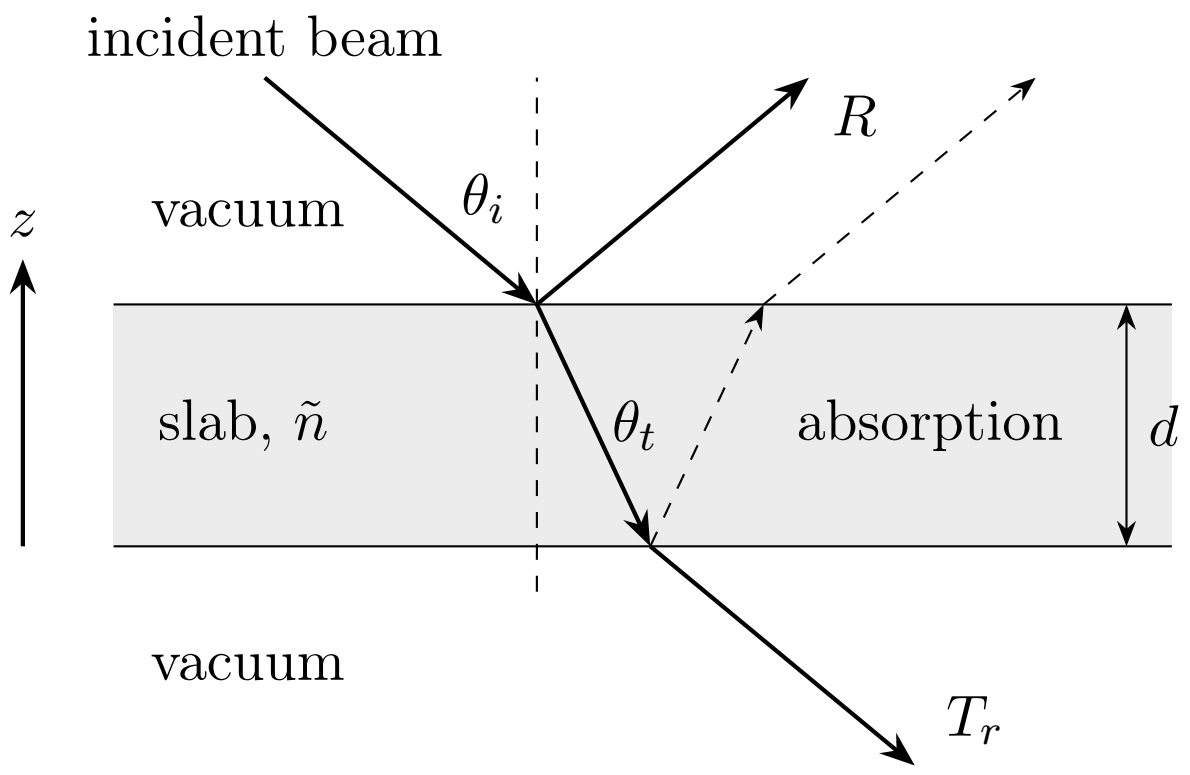

Sunlight is incoherent. The phase between multiply reflected beams is not maintained, so interference between internal reflections can be neglected. We therefore sum the reflections at intensity level (incoherent summation). Each material is treated as a plane-parallel slab of thickness $d$ with vacuum on both sides \cite{Polyanskiy2024-cv}. Figure~\ref{fig:slab} shows the geometry. The $z$ axis is the thickness direction, normal to the two surfaces.

Inside the slab, the wave has the wave-vector component along $z$ \cite{Reider2016-ic}
\begin{equation}
k_z=\frac{2\pi}{\lambda}\sqrt{\tilde n^2-\sin^2\theta_i}
\end{equation}
The imaginary part of $k_z$ describes absorption. The single-pass internal transmittance for thickness $d$ is \cite{Polyanskiy2024-cv}
\begin{equation}
\tau=\exp\left\{-2\,\mathrm{Im}[k_z]\,d\right\}
\end{equation}

The beam crosses two interfaces and is attenuated inside the slab. Repeated internal reflections form a geometric series. For each polarization $\omega$, slab transmissivity and reflectivity are \cite{Reider2016-ic}

\begin{equation}
T_{r,\omega}=\frac{(1-R_{\mathrm{F},\omega})^2\,\tau}{1-R_{\mathrm{F},\omega}^{2}\tau^2},
\qquad
R_{\omega}=R_{\mathrm{F},\omega}+R_{\mathrm{F},\omega}\,\frac{(1-R_{\mathrm{F},\omega})^2\,\tau^2}{1-R_{\mathrm{F},\omega}^{2}\tau^2}
\end{equation}
In $R_\omega$, the first term is the reflection at the first interface. The second term accounts for the light that enters the slab, is reflected internally, and leaves through the illuminated surface. The unpolarized slab properties are the averages over $s$ and $p$, and the absorptivity follows from energy conservation \cite{BornWolf2019}:

\begin{equation}
R_\mathrm{dir}=\tfrac{1}{2}(R_s+R_p),
\quad
T_{r,\mathrm{dir}}=\tfrac{1}{2}(T_{r,s}+T_{r,p}),
\quad
A_\mathrm{dir}=1-R_\mathrm{dir}-T_{r,\mathrm{dir}}
\end{equation}

\paragraph{Numerical safeguards}
The denominator $1-R_{\mathrm{F},\omega}^{2}\tau^2$ comes from the sum of the geometric series of internal reflections. The closed form is exact, so no term is cut off. The denominator is zero only if $R_{\mathrm{F},\omega}=1$ and $\tau=1$ at the same time. This means a lossless slab at grazing incidence ($90^\circ$). The largest angle of the grid is $89.9^\circ$ ($90^\circ$ is not used). Even for a lossless slab with a refractive index up to 4, the denominator stays above $3\times10^{-3}$ at this angle. For absorbing materials, $\tau<1$ and the denominator is larger. For metals, $\tau\approx0$ and the denominator is close to 1. As a numerical guard against round-off, the denominator is kept above $10^{-30}$. However, this guard was never triggered, in practice.

\subsubsection{Hemispherical properties and calculation grids}
\label{sec:grids}

The hemispherical property is the cosine-weighted average of the directional property over the hemisphere:
\begin{equation}
X_\mathrm{hemi}=\frac{1}{\pi}\int_0^{2\pi}\!\!\int_0^{\pi/2}X_\mathrm{dir}\cos\theta_i\sin\theta_i\,d\theta_i\,d\varphi
=2\int_0^{\pi/2}X_\mathrm{dir}\cos\theta_i\sin\theta_i\,d\theta_i
\label{eq:hemi}
\end{equation}
with $X=R$, $T_r$, or $A$. This is a full three-dimensional average over the solid angle $d\Omega=\sin\theta_i\,d\theta_i\,d\varphi$. Because the slab is isotropic, $X_\mathrm{dir}$ does not depend on the azimuth angle $\varphi$. The $\varphi$ integral then gives $2\pi$, and the normalization $\int_0^{2\pi}\!\int_0^{\pi/2}\cos\theta_i\sin\theta_i\,d\theta_i\,d\varphi=\pi$ gives the factor 2. We evaluate the integral with the trapezoidal rule.

The database gives the slab properties for the following grids:

\begin{itemize}
\item incidence angle $\theta_i=0^\circ,5^\circ,10^\circ,\ldots,85^\circ,89.9^\circ$
\item thickness $d=0.001,\,0.01,\,0.1,\,1,\,10,\,100,\,1000,\,10000~\mu$m
\item temperature: a material-specific grid of 50 points from $T_\mathrm{low}$ to $T_\mathrm{up}$
\end{itemize}

$T_\mathrm{low}$ is the lowest available temperature of the thermal conductivity data of the material (Section~\ref{sec:kappa}). It lies between 5~K and 373~K. $T_\mathrm{up}$ is the upper temperature limit (Section~\ref{sec:upperlimit}). The optical properties are given as three-dimensional arrays (wavelength, angle, thickness). For sources with temperature-dependent data, temperature is a fourth dimension. For ``-Tdepend'' entries, $A$, $R$, and $T_r$ are interpolated linearly in temperature between the temperatures of the source.

\subsubsection{Upper temperature limit}
\label{sec:upperlimit}

The source data and the slab model describe the solid state. We therefore limit all results to a material-specific \emph{upper temperature limit} $T_\mathrm{up}$. It is the highest temperature up to which the refractive-index data and the models are considered valid. For most materials, $T_\mathrm{up}$ is the melting point. Specifically, \ce{Ge}: 1210.6~K, \ce{Si}: 1685.2~K, \ce{SiO2}: 1986.0~K, \ce{TiC}: 3523.0~K, \ce{ZrO2}: 2973.0~K, \ce{Ag}: 1234.0~K, \ce{Al}: 933.2~K, \ce{Al2O3}: 2323.0~K, \ce{Au}: 1336.2~K, \ce{Cu}: 1356.0~K, \ce{Fe}: 1810.0~K, \ce{Ni}: 1726.0~K, \ce{Ti}: 1953.0~K, \ce{TiO2}: 2133.0~K. Above the melting point, the material is liquid, and its optical properties cannot be derived from solid-state data. For seven materials, there is no normal melting point or another process happens at a lower temperature, above which either the literature data or the models are not valid. Table~\ref{tab:upperlimit} lists these cases. $T_\mathrm{up}$ is stored for every material (field \fld{upper_temperature_limit_K}) and is shown in the App. No property is calculated above $T_\mathrm{up}$.

\begin{table}[t]
\centering
\caption{Materials whose upper temperature limit $T_\mathrm{up}$ is not the melting point.}
\label{tab:upperlimit}
\begin{tabular}{>{\centering\arraybackslash}m{0.27\textwidth}
>{\centering\arraybackslash}m{0.23\textwidth}
>{\centering\arraybackslash}m{0.4\textwidth}}
\hline
\textbf{Material} & \textbf{$T_\mathrm{up}$ (K)} & \textbf{Limiting process} \\
\hline
Amorphous carbon & 3925 & sublimation \\
PMMA & 332.2 & softening \\
\ce{SiC} & 3100 & decomposition into molten or vaporized Si and solid carbon \\
\ce{Cu2O} & 423 & reduction to Cu \\
\ce{Fe2O3} & 950 & N\'eel temperature (loss of magnetic order) \\
\ce{Fe3O4} & 858 & Curie temperature (loss of magnetic order) \\
Kapton-type polyimide & 873.2 & volatilization and carbonization \\
\hline
\end{tabular}
\end{table}

\subsection{Interaction with solar radiation}
\label{sec:solar}

\begin{figure}[t]
    \centering
    \includegraphics[width=1\linewidth]{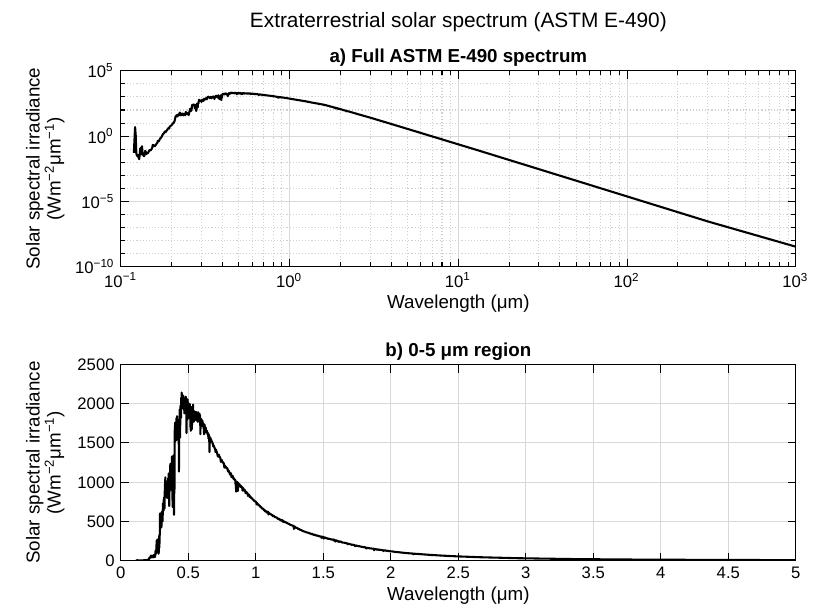}
    \caption{Extraterrestrial solar spectrum of the ASTM E490 standard: (a) full spectrum, (b) zoom on the $0$--$5~\mu$m range, where the peak irradiance is found.}
    \label{fig:am0}
\end{figure}

The database is meant for sunlight in space. We therefore use the extraterrestrial Air Mass Zero (AM0) spectrum of the ASTM E490-00 standard \cite{ASTM_E490}. It gives the spectral irradiance $I_\mathrm{solar}(\lambda)$ outside the Earth's atmosphere (Figure~\ref{fig:am0}). The irradiance is defined on a surface normal to the beam.

The AM0 wavelength grid differs from the grids of the materials' sources. We use the AM0 grid as the master grid. The optical constants of each source are resampled on it by linear interpolation before any calculation. From here on, $\lambda_s$ is a wavelength on this grid. A source may not cover the whole AM0 range, so all calculations are limited to the common interval of the AM0 and the literature source spectrum
\begin{equation}
\lambda_{s,\min}=\max\left(\lambda_{\min}^{\mathrm{source}},\lambda_{\min}^{\mathrm{AM0}}\right),
\qquad
\lambda_{s,\max}=\min\left(\lambda_{\max}^{\mathrm{source}},\lambda_{\max}^{\mathrm{AM0}}\right)
\label{eq:lambda_limits}
\end{equation}

The spectral absorbed, reflected, and transmitted irradiances are the optical properties weighted with the AM0 spectrum:
\begin{equation}
\begin{aligned}
q_{\mathrm{abs},\lambda_s}&=A_\mathrm{dir}\,I_{\mathrm{solar}}(\lambda_s),\\
q_{\mathrm{ref},\lambda_s}&=R_\mathrm{dir}\,I_{\mathrm{solar}}(\lambda_s),\\
q_{\mathrm{tr},\lambda_s}&=T_{r,\mathrm{dir}}\,I_{\mathrm{solar}}(\lambda_s)
\end{aligned}
\end{equation}
The directional solar fluxes are their integrals over wavelength (trapezoidal rule \cite{Kythe_2004-nw}), with $\mathrm{Y}=\mathrm{abs},\mathrm{ref},\mathrm{tr}$, denoting for ``absorbed'', ``reflected'' and ``transmitted'', respectively:
\begin{equation}
Q_{\mathrm{Y},\mathrm{dir}}(T^\ast,\theta_i,d)=\int_{\lambda_{s,\min}}^{\lambda_{s,\max}}q_{\mathrm{Y},\lambda_s}\,d\lambda_s
\label{eq:Q_dir}
\end{equation}
$Q_{\mathrm{Y},\mathrm{dir}}$ is the directional (absorbed, reflected, transmitted) solar flux. The hemispherical fluxes are
\begin{equation}
Q_{\mathrm{Y},\mathrm{hemi}}(T^\ast,d)=2\int_0^{\pi/2}Q_{\mathrm{Y},\mathrm{dir}}\cos\theta_i\sin\theta_i\,d\theta_i
\end{equation}
Here, ``hemispherical'' means an average over the incidence direction of one collimated solar beam. It does not describe radiation from a spatially diffuse source.

Each calculation is repeated for every combination of angle, thickness, and temperature. The AM0 spectrum depends only on wavelength, so any change of the fluxes comes only from the optical properties.

\paragraph{Concentrated sunlight}
Mirrors or lenses can focus sunlight onto a small area. The solar concentration $C$ is the ratio between the irradiance on the target and the AM0 solar irradiance, which is about $1.366$~kW\,m$^{-2}$ and is called ``1 sun''. For example, $C=1000$ suns means about $1.366$~MW\,m$^{-2}$. We scale the fluxes linearly:
\begin{equation}
Q^{\mathrm{conc}}_{\mathrm{Y},\mathrm{dir}}=C\,Q_{\mathrm{Y},\mathrm{dir}},
\qquad
Q^{\mathrm{conc}}_{\mathrm{Y},\mathrm{hemi}}=C\,Q_{\mathrm{Y},\mathrm{hemi}}
\end{equation}
This assumes an ideal optical system. The spectrum keeps the AM0 shape, and the beam is treated as collimated. The database gives the fluxes for $C=1$, 2, 4, 6, \ldots, 20 (steps of 2), 30, 40, \ldots, 100 (steps of 10), 200, 300, 400, 500, 1000, and 2000 suns. A concentration range between 1 and 2000 suns is typical for solar concentrating systems and has also been adopted in previous studies \cite{Maka2020-hb}.

\subsection{Emissivity}
\label{sec:emissivity}

To find the thermal response, we also need to know the ability of a material to emit radiation. Instead of assuming a blackbody or a gray body, we calculate the wavelength-dependent emissivity from the same optical data given by the literature sources. Kirchhoff's law \cite{Mobedi2023-qo} states that, at thermal equilibrium, the directional spectral emissivity equals the spectral absorptivity at the same wavelength, temperature, angle, and thickness:
\begin{equation}
\varepsilon_{\lambda,\mathrm{dir}}(\lambda_s,T^\ast,\theta_i,d)=A_\mathrm{dir}(\lambda_s,T^\ast,\theta_i,d)
\end{equation}
For emissivity, $\theta_i$ denotes the emission angle and $A_{dir}$ is the directional absorptivity, calculated in Section~\ref{sec:optics}.

The radiance of a real body is the blackbody radiance multiplied by the directional spectral emissivity \cite{Naterer2021-cn}, and the blackbody radiance follows Planck's law:
\begin{equation}
B_{\lambda,\mathrm{real}}=\varepsilon_{\lambda,\mathrm{dir}}\,B_{\lambda,\mathrm{bb}},
\qquad
B_{\lambda,\mathrm{bb}}(\lambda_s,T)=\frac{2hc^2}{\lambda_s^5}\left[\exp\left(\frac{hc}{\lambda_s k_B T}\right)-1\right]^{-1}
\label{eq:planck}
\end{equation}
Planck's law is dimensionally consistent only if $\lambda_s$ is in metres. All other equations use micrometres. The code converts $\lambda_s$ to metres before it evaluates Planck's law.

The emissivity depends on temperature for all materials, because the Planck weighting changes with temperature. For sources with temperature-dependent optical data, $\varepsilon_{\lambda,\mathrm{dir}}$ also changes with temperature. The total directional emissivity is \cite{Cengel2003-cn}
\begin{equation}
\varepsilon_\mathrm{dir}(T,\theta_i,d)=
\frac{\displaystyle\int_{\lambda_{s,\min}}^{\lambda_{s,\max}}\varepsilon_{\lambda,\mathrm{dir}}\,B_{\lambda,\mathrm{bb}}\,d\lambda_s}
{\displaystyle\int_{\lambda_{s,\min}}^{\lambda_{s,\max}}B_{\lambda,\mathrm{bb}}\,d\lambda_s}
\label{eq:epsilon_dir}
\end{equation}
Both integrals use only the wavelength range covered by the optical data of the literature source, $[\lambda_{s,\min},\lambda_{s,\max}]$, and not the full Planck spectrum (0 to $\infty$). The emissivity is therefore normalized inside this spectral window. At low temperatures, the emission peak lies at long wavelengths, for which some literature sources do not provide optical data. The emissivity is then poorly constrained. To quantify this, we introduce in Section~\ref{sec:resampled_check} the thermal coverage ratio $\eta_\mathrm{th}$ (Eq.~\ref{eq:thermal_coverage_ratio}). It is the fraction of the Planck spectrum inside the spectral window, and it is stored for every source and temperature. It shows for which temperatures each source gives a reliable emissivity: the closer $\eta_\mathrm{th}$ is to 1, the better. 

The hemispherical emissivity and the emitted flux are
\begin{equation}
\varepsilon_\mathrm{hemi}(T,d)=2\int_0^{\pi/2}\varepsilon_\mathrm{dir}\cos\theta_i\sin\theta_i\,d\theta_i,
\qquad
Q_{\mathrm{emit},\mathrm{hemi}}(T,d)=\varepsilon_\mathrm{hemi}\,\sigma_B\,T^4
\label{eq:Qemit}
\end{equation}
where the second expression is given by the Stefan--Boltzmann law \cite{Mobedi2023-qo}.

It is noted that the directional emissivity $\varepsilon_{dir}(T,\theta_i,d)$, calculated from Eq.~\ref{eq:epsilon_dir}, is retained in the database as a fundamental temperature, angle- and thickness-dependent material property. Such directional properties are central to directional radiative heat-transfer analyses, where the exchanged power depends on the relative positions, orientations, and separation of the emitting and receiving surfaces \cite{Pan2026-su}. In the present database, however, we consider an isolated material slab (e.g., a debris fragment) emitting into the outward hemisphere (e.g., into the space), rather than a specific emitter--receiver configuration. Therefore, a standalone directional emitted flux is not given in our database.

\subsection{Equilibrium temperature}
\label{sec:temperature}

The temperature of a material exposed to solar irradiation is an important quantity that determines its thermal response, as the emitted thermal
radiation depends strongly on temperature. We estimate it with two models. Both assume that (i) convection is negligible, (ii) radiative heat exchange dominates the energy balance (e.g., in space), and (iii) the concentrated solar flux of Section~\ref{sec:solar} is the only heat input. For the reasons explained in Section~\ref{sec:emissivity}, both models use the hemispherical emitted flux of Eq.~(\ref{eq:Qemit}). Absorption can be directional or hemispherical.

For a flat surface, such as the one in Figure~\ref{fig:slab}, tilted by $\theta_i$, the power per unit surface area is $\cos\theta_i\,Q_{\mathrm{abs},\mathrm{dir}}$, because $Q_{\mathrm{abs},\mathrm{dir}}$ refers to the beam cross-section (Eq.~\ref{eq:Q_dir}). We use this projected flux in all energy balances. Both models are solved on the material-specific temperature grid of Section~\ref{sec:grids}.

\subsubsection{Lumped-body model}
\label{sec:lumped}

The body is treated as one lumped thermal mass with uniform temperature. In steady state, the absorbed flux equals the emitted flux \cite{Cengel2003-cn}. The equilibrium temperature $T_\mathrm{eq}$ is the root of one of two residual functions \cite{Naterer2021-cn}:
\begin{equation}
\begin{aligned}
\Delta Q_\mathrm{dir\text{-}hemi}(T_\mathrm{eq},\theta_i,d)&=Q_{\mathrm{emit},\mathrm{hemi}}(T_\mathrm{eq},d)-\cos\theta_i\,Q^{\mathrm{conc}}_{\mathrm{abs},\mathrm{dir}}(T_\mathrm{eq},\theta_i,d)=0,\\
\Delta Q_\mathrm{hemi\text{-}hemi}(T_\mathrm{eq},d)&=Q_{\mathrm{emit},\mathrm{hemi}}(T_\mathrm{eq},d)-Q^{\mathrm{conc}}_{\mathrm{abs},\mathrm{hemi}}(T_\mathrm{eq},d)=0
\end{aligned}
\end{equation}
The first line is directional absorption with hemispherical emission. The second line is hemispherical absorption with hemispherical emission. The root is the first sign change of $\Delta Q$ on the temperature grid, refined by linear interpolation between the two neighbouring grid points. If there is no sign change inside $[T_\mathrm{low},T_\mathrm{up}]$, no equilibrium exists in this range. This usually means that the equilibrium lies above $T_\mathrm{up}$, so the material melts or degrades under these conditions. For materials whose grid starts at a high temperature, it can also mean an equilibrium below $T_\mathrm{low}$, for example at low solar concentration. Since our models calculate results only inside $[T_\mathrm{low},T_\mathrm{up}]$, we make no assumptions outside it, such as extrapolation or constant values. Instead, we store NaN in the corresponding cells of the database. This informs the user that no equilibrium exists in the calculated range.

The lumped-body energy balance has one emitting surface per illuminated surface. It describes, for example, a plate with an insulated rear face. A free-standing plate emits from all faces, which the following slab model includes.

\subsubsection{Slab model with heat conduction}
\label{sec:slab}

The lumped model assumes that the temperature in the body is uniform. This holds for thin or highly conductive bodies, but not in general. The slab model therefore adds heat conduction through the thickness. We consider a free-standing, homogeneous, plane-parallel slab of thickness $d$ and lateral size $L\gg d$, under the following conditions:
\begin{itemize}
\item the absorbed solar power is uniformly deposited at the front (illuminated) surface
\item the lateral extent is large, so that edge effects are neglected: the side faces have an area fraction of order $d/L$ and are not modeled, and conduction is steady and one-dimensional along the thickness, with temperature-dependent thermal conductivity $\kappa(T)$ and no internal heat source
\item both faces (front and rear) radiate hemispherically into cold space, each with the emissivity $\varepsilon_\mathrm{hemi}$ at its own temperature.
\end{itemize}

Fourier's law, integrated over the thickness, gives the conductive flux from the front temperature $T_\mathrm{f}$ to the rear temperature $T_\mathrm{b}$:
\begin{equation}
\label{eq:conduction}
q_\mathrm{cond}=\frac{1}{d}\int_{T_\mathrm{b}}^{T_\mathrm{f}}\kappa(T)\,dT
\end{equation}
The energy balances at the two surfaces are
\begin{equation}
\label{eq:slab_balance}
\begin{aligned}
Q_\mathrm{abs}^{\mathrm{conc}}&=\varepsilon_\mathrm{hemi}(T_\mathrm{f})\,\sigma_B T_\mathrm{f}^4+q_\mathrm{cond}\quad\text{(front)},\\
q_\mathrm{cond}&=\varepsilon_\mathrm{hemi}(T_\mathrm{b})\,\sigma_B T_\mathrm{b}^4\quad\text{(rear)}
\end{aligned}
\end{equation}
$Q_\mathrm{abs}^{\mathrm{conc}}$ can be either the projected directional flux or the hemispherical flux, as in Section~\ref{sec:lumped}. The two equations give $T_\mathrm{f}$ and $T_\mathrm{b}$. For steady conduction without heat sources, $\frac{\partial q_{cond}}{\partial z}=0$, thus the integral $\int\kappa(T)\,dT$ depends linearly on thickness. The mid-plane temperature $T_\mathrm{mid}$ is therefore defined by $\int_{T_\mathrm{b}}^{T_\mathrm{mid}}\kappa(T)\,dT=\tfrac{1}{2}\int_{T_\mathrm{b}}^{T_\mathrm{f}}\kappa(T)\,dT$.

Eqs.~\ref{eq:conduction} and~\ref{eq:slab_balance} are solved iteratively. The initial value trialed is the isothermal solution ($T_\mathrm{f}=T_\mathrm{b}$), which solves $Q_\mathrm{abs}^{\mathrm{conc}}=2\varepsilon_\mathrm{hemi}\sigma_B T^4$ on the temperature grid. If the solver does not converge, or finds no difference between front and rear, the isothermal solution is returned. This is the correct result when the conduction is negligible, for example for thin or highly conductive slabs. If no isothermal solution exists inside $[T_\mathrm{low},T_\mathrm{up}]$, NaN is inserted in the respective cells of the database, as discussed in Section~\ref{sec:lumped}.

\paragraph{Thermal conductivity data}
\label{sec:kappa}
The temperature-dependent thermal conductivities $\kappa(T)$ of the materials, included in the database, come from the recommended values of Volumes 1 and 2 of the TPRC thermal conductivity series \cite{TouloukianVol1,TouloukianVol2}. For Kapton we use the data from ref.~\cite{Venkatachalam2017}. The tabulated $(T,\kappa)$ pairs and their references are stored in the database.

\paragraph{Sources measured with a substrate}

Most literature sources report optical data for a single material, either bulk or thin film. A few sources, however, report optical constants that describe the whole system of a thin film, its substrate, and possibly an adhesion (wetting) layer. These are the studies of Ciesielski et al. on Ag thin films deposited on $\mathrm{SiO_2}$ substrates, either directly or with a Ge or Ni adhesion layer \cite{Ciesielski2017-jj}, and on Au thin films deposited on $\mathrm{SiO_2}$, either directly or with a Ge adhesion layer \cite{Ciesielski2018-Au}. The thermal conductivities in the database are those of the pure material, so conduction results would not be meaningful for these sources. For these silver and gold sources, we therefore set the conductive flux and the slab temperatures to NaN to inform the user that no slab temperatures were estimated. The optical and radiative quantities and the lumped temperatures do not depend on $\kappa$, so they are kept.

\section{Data Record}
\label{sec:datarecord}

The database presented here, SpaderOptics, contains the optical, solar-interaction, thermal-emission, equilibrium-temperature and slab-temperature data generated for the investigation of concentrated solar radiation interactions with different metallic and non-metallic materials. The database integrates refractive indices and extinction coefficients obtained from published literature \cite{Hagemann1975-sg,Arakawa1977-pb,Larruquert2013-nx,Amotchkina2020-dh,Aspnes1983-gr,Burnett2016-vx, Barnes1979-hy,Icenogle1976-cy,Jellison1992-et,Li1980-nj,Nunley2016-fr,Ciesielski2018-as,Arakawa1981-im,Brannon1985-wb,Cunningham2011-js,French2009-dv,Kawka2001-zd,Kumar2011,Philipp1986-fq,Sahin2019-ci,Smith1975-og,Zhang1998-hy,Beadie2015-ka,Bodurov2016-ka,Nyakuchena2023-wn,Sultanova2009-bl,Szczurowski2013,Tsuda2018-gg,Zhang2020-vr,Chandler-Horowitz2005-ei,Schinke2013-ud,Edwards1980-in,Green1995-rt,Green2008-mu,Salzberg1957-sl,Schinke2015-py,Shkondin2017-ym,Sik1998-sx,Vuye1993-se,Wang2025-cv,Franta2013-qz,Karaman2025-ua,Pierce1972-rc,Shaffer1971-za,Fischer2017-yw,Wang2013-mn,Singh1971-fc,Larruquert2011-av,Ghosh1999-cs,Radhakrishnan1951-gd,Arosa2020-ny,Franta2016-jo,Malitson1965-sc,Kitamura2007-qf,Gao2013-yj,Kischkat2012-td,Gao2012-cp,Rodriguez-de_Marcos2016-pw,Pfluger1984-wo,Synowicki2004-fq,Wood1982-gz,Babar2015-ir,Choi2020-nm,Ciesielski2017-jj,Jiang2016-zc,Johnson1972-fh,Stahrenberg2001-nk,Werner2009-me,Windt1988-tn,Wu2014-nb,Yang2015-gt,Boidin2016-sz,Franta2015-li,Malitson1962-ub,Zhukovsky2015-cc,Klinavicius2025-jx,Gao2011-bk,Magnozzi2019-bh,Olmon2012-yf,Ordal1987-bs,Rosenblatt2020-pt,Yakubovsky2017-ou,Ordal1985-zb,Mash1973-zb,Palm2018-ed,Bond1965-vl,DeVore1951-tw,Jolivet2023-ds,Sarkar2019-on,Siefke2016-iq,Ciesielski2018-Au} with derived optical properties, solar flux quantities, thermal-emission quantities, equilibrium and slab temperatures calculated under the conditions described in the Methods section.

The complete database is publicly available through Zenodo at \fld{https://doi.org/10.5281/zenodo.23643910}~\cite{spaderoptics_data} under a CC-BY-4.0 licence. The deposited dataset is organized in a MATLAB structure, is provided as a MATLAB .mat file, called SpaderOptics.mat, and uses the MATLAB -v7.3 format. 

The database is organized hierarchically by material and literature source. Each \fld{material} entry contains a \fld{dataset} structure comprising one or more literature sources, and a \fld{source_spread} structure that quantifies how much the calculated hemispherical emissivity varies from one literature source to another (see Section~\ref{sec:spread}). Each source record contains the bibliographic information, original refractive indices and extinction coefficients (as given in the YAML files, provided in the RefractiveIndex.INFO database \cite{Polyanskiy2024-cv}), thermal conductivity data from literature \cite{TouloukianVol1,TouloukianVol2,Venkatachalam2017}, resampled optical constants, calculated optical properties, solar fluxes, thermal-emission quantities, equilibrium temperatures, slab temperatures and spectral-coverage metrics associated with that source.

For materials for which the literature reports $n$ and $k$ as functions of both wavelength and temperature, the database preserves the optical-property data corresponding to each reported temperature. These datasets are subsequently combined into a new temperature-dependent entry, denoted by the corresponding author name followed by "-Tdepend".

\begin{figure}[H]
	\centering    \includegraphics[width=0.65\linewidth]{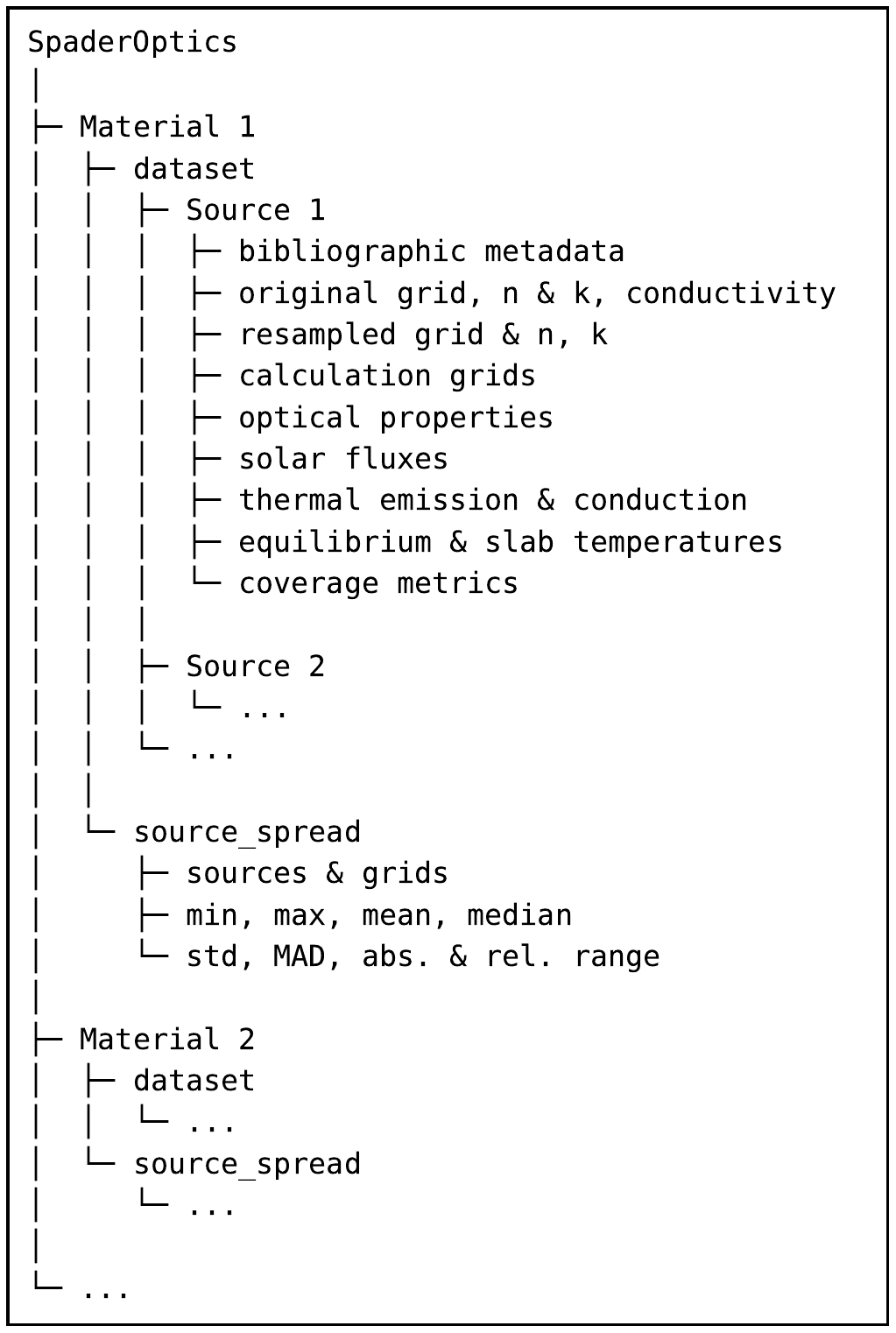}
	\caption{Schematic representation of the structure and organization of the SpaderOptics database.}
	\label{fig:Database_structure}
\end{figure}

Figure~\ref{fig:Database_structure} provides a schematic overview of the organization of the SpaderOptics database and the relationship between the original literature data, resampled optical constants, and the derived optical, solar-flux, thermal-emission, conductive-flux, equilibrium-temperature and slab-temperature quantities, as well as the source-to-source spread of the calculated hemispherical emissivity.

In each \fld{dataset} struct and source record the following groups of information are contained: 

\begin{itemize}
	\item The fields:
	\begin{itemize}
		\item \fld{author},
		\item \fld{reference}, and
		\item \fld{source_url}
	\end{itemize} 
	
	identify the literature source from which the original optical constants were obtained.
	
	\item The field \fld{upper_temperature_limit_K} (in [K]) gives the highest temperature up to which the database contains data for each material. This limit is the melting point, or the decomposition or other transition temperature of the material, as listed in Table~\ref{tab:upperlimit}.
	
	\item The fields:
	\begin{itemize}
		\item \fld{thermal_conductivity_temperature_grid_K} (in [K]),
		\item \fld{thermal_conductivity_W_mK} (in $[\mathrm{Wm^{-1}K^{-1}}]$), and
		\item \fld{thermal_conductivity_reference}
	\end{itemize} 
	
	give data relevant to the thermal conductivity $\kappa(T)$ of each material. The first two store the tabulated temperatures, and the corresponding values of $\kappa$. The third identifies the literature source of these data. These data are used to solve the slab-temperature model in Section~\ref{sec:slab}.
	
	\item The fields 
	\begin{itemize}
		\item \fld{original_temperature_K} (in[K]), and
		\item \fld{original_thickness_um} (in $[\mathrm{\mu m}]$)
	\end{itemize} 
	
	describe the temperature and thickness, if available, associated with the experimental conditions used by the respective authors to obtain the original optical dataset. If this information is not available by the source then those fields have a $('')$ instead of a value.
	
	\item The fields:
	\begin{itemize}
		\item \fld{original_wavelength_um} (in $[\mathrm{\mu m}]$), and
		\item \fld{original_n_complex} 
	\end{itemize} 
	
	contain the wavelength-dependent complex refractive indices as obtained from the literature, before resampling onto the solar wavelength grid used for the calculations.
	
	\item The fields:
	\begin{itemize}
		\item \fld{resampled_wavelength_um} (in $[\mathrm{\mu m}]$), and
		\item \fld{resampled_n_complex}
	\end{itemize} 
	
	contain the wavelength-dependent complex refractive indices after interpolation onto the solar wavelength grid used for the calculations of all the quantities that follow.
	
	\item The fields:
	\begin{itemize}
		\item \fld{calculation_incident_angles_deg} (in [deg]),
		\item \fld{calculation_thickness_um} (in $[\mathrm{\mu m}]$),
		\item \fld{calculation_temperature_K} (in [K]), and
		\item \fld{calculation_solar_concentration_multipliers}
	\end{itemize}
	
	define the grids of incident/emission angle, material thickness, temperature, and solar concentration multipliers over which the derived quantities are calculated.
	
	\item The fields:
	\begin{itemize}
		\item \fld{calculated_R_directional},
		\item \fld{calculated_T_directional},
		\item \fld{calculated_A_directional}, 
		\item \fld{calculated_E_directional}, 
		\item \fld{calculated_R_hemispherical}, 
		\item \fld{calculated_T_hemispherical}, 
		\item \fld{calculated_A_hemispherical}, and
		\item \fld{calculated_E_hemispherical}
	\end{itemize}
	
	contain the calculated directional and hemispherical reflectivity, transmissivity, absorptivity, and emissivity.
	
	For the directional reflectivity, transmissivity, and absorptivity arrays, the dimensions are ordered as $\lambda_s \times \theta_i \times d$. For temperature-dependent source datasets, an additional fourth dimension is included, ordered as $\lambda_s \times \theta_i \times d \times T$. The directional emissivity array is ordered as temperature $T \times \theta_i \times d$.
	
	The hemispherical reflectivity, transmissivity, and absorptivity arrays are ordered as $\lambda_s \times d$. For temperature-dependent source datasets, the stored dimensions are $\lambda_s \times d \times T$. Hemispherical emissivity is stored as temperature $T \times d$.
	
	\item The fields:
	\begin{itemize}
		\item \fld{calculated_directional_Reflected_flux_W_m2} (in $[\mathrm{Wm^{-2}}]$), 
		\item \fld{calculated_directional_Transmitted_flux_W_m2} (in $[\mathrm{Wm^{-2}}]$),  
		\item \fld{calculated_directional_Absorbed_flux_W_m2} (in $[\mathrm{Wm^{-2}}]$), 
		\item \fld{calculated_hemispherical_Reflected_flux_W_m2} (in $[\mathrm{Wm^{-2}}]$), 
		\item \fld{calculated_hemispherical_Transmitted_flux_W_m2} (in $[\mathrm{Wm^{-2}}]$), 
		\item \fld{calculated_hemispherical_Absorbed_flux_W_m2} (in $[\mathrm{Wm^{-2}}]$)
	\end{itemize}
	
	contain the directional and hemispherical reflected, transmitted, and absorbed solar fluxes.
	
	The directional reflected, transmitted, and absorbed solar flux arrays are ordered as $\theta_i \times d$. For temperature-dependent source datasets, the temperature dimension is included, giving the ordering $T \times \theta_i \times d$. 
	
	The hemispherical reflected, transmitted, and absorbed flux arrays depend only on the material thickness for non-temperature-dependent sources, and they are ordered as $T \times d$ for temperature-dependent sources.
	
	\item The field \fld{calculated_hemispherical_Emitted_flux_W_m2} (in $[\mathrm{Wm^{-2}}]$) contains the hemispherical emitted flux. The hemispherical emitted flux array is ordered as $T \times d$.
	
	\item The fields:
	\begin{itemize}
		\item \fld{calculated_lumped_equilibrium_temperature_dirA_hemiE_K} 
		
		(in [K]), and 
		\item \fld{calculated_lumped_equilibrium_temperature_hemiA_hemiE_K} 
		
		(in [K])
	\end{itemize}
	
	contain equilibrium temperatures calculated using the alternative combinations of indicated directional or hemispherical absorptivity and hemispherical emissivity.
	
	The first array is ordered as $\theta_i \times d \times C$, whereas the second array is ordered as $d \times C$.
	
	\item The fields:
	\begin{itemize}
		\item \fld{calculated_Conductive_flux_dirA_hemiE_W_m2} (in [W\,m$^{-2}$]), and
		\item \fld{calculated_Conductive_flux_hemiA_hemiE_W_m2} (in [W\,m$^{-2}$])
	\end{itemize}
	
	contain the steady-state conductive heat flux $q_\mathrm{cond}$ through the slab, from the front to the rear face, for the two combinations of directional or hemispherical absorptivity and hemispherical emissivity. The flux is positive when heat flows from the illuminated face to the rear face.
	
	The first array is ordered as $\theta_i \times d \times C$, whereas the second array is ordered as $d \times C$.
	
	\item The fields:
	\begin{itemize}
		\item \fld{calculated_slab_equilibrium_temperature_front_dirA_hemiE_K} 
		
		(in [K]),
		\item \fld{calculated_slab_equilibrium_temperature_rear_dirA_hemiE_K} 
		
		(in [K]), 
		\item \fld{calculated_slab_equilibrium_temperature_mid_dirA_hemiE_K} 
		
		(in [K]) 
	
		\bigskip and \bigskip
		
		\item \fld{calculated_slab_equilibrium_temperature_front_hemiA_hemiE_K} (in [K]), 
		\item \fld{calculated_slab_equilibrium_temperature_rear_hemiA_hemiE_K} (in [K]),
		\item \fld{calculated_slab_equilibrium_temperature_mid_hemiA_hemiE_K} (in [K]) 
	\end{itemize}
	
	contain the steady-state temperatures of the slab model (Section~\ref{sec:slab}) at the front (illuminated) face, $T_\mathrm{f}$, at the rear face, $T_\mathrm{b}$, and at the mid-plane, $T_\mathrm{mid}$. The first group uses directional absorptivity and the second group uses hemispherical absorptivity, both combined with hemispherical emissivity.
	
	The arrays of the first group are ordered as $\theta_i \times d \times C$, whereas the arrays of the second group are ordered as $d \times C$.
	
	\item The fields:
	\begin{itemize}
		\item \fld{metrics_solar_coverage_ratio}, and 
		\item \fld{metrics_thermal_coverage_ratio}
	\end{itemize}
	
	provide spectral-coverage metrics associated with the underlying optical datasets. 
	
	The solar coverage ratio represents the fraction of the AM0 solar spectral irradiance covered by the wavelength interval available for a given literature source. This is a scalar quantity and depends only on the literature source.
	
	The thermal coverage ratio describes the corresponding wavelength coverage relevant to the thermal emission calculations. It is a vector ordered according to $T$.
	
\end{itemize}

In addition to the \fld{dataset} structure, each \fld{material} entry contains a \fld{source_spread} structure, which quantifies the variability between the different literature sources of that material in terms of the calculated hemispherical emissivity. Specifically, it contains the following groups of information:

\begin{itemize}
	\item The fields:
	\begin{itemize}
		\item \fld{description},
		\item \fld{authors},
		\item \fld{temperature_K} (in [K]), and
		\item \fld{thickness_um} (in $[\mathrm{\mu m}]$)
	\end{itemize}
	
	provide a short description of the structure, the list of literature sources included in the spread, and the temperature and thickness grids over which the spread is defined.
	
	\item The fields:
		\begin{itemize}
		\item \fld{min},
		\item \fld{max},
		\item \fld{mean},
		\item \fld{median},
		\item \fld{std},
		\item \fld{mad},
		\item \fld{range_abs}, and
		\item \fld{range_rel_pct} (in [\%])
	\end{itemize}
	
	contain, respectively, the minimum, maximum, mean, median, standard deviation (std), median absolute deviation (MAD), absolute range and relative range (in percent of the median) of the hemispherical emissivity across the included sources. All of them are ordered as temperature $T \times d$, the same ordering as \fld{calculated_E_hemispherical}.
\end{itemize}

The spread is available only for materials that include at least two sources; otherwise \fld{source_spread} is empty.

Overall, the deposited database preserves the relationship between each derived quantity and its corresponding material, literature source, wavelength grid, and calculation conditions. Consequently, individual records can be traced from the original literature optical constants through the resampled optical data to the calculated optical, solar-flux, thermal-emission, equilibrium-temperature, and slab-temperature quantities, as well as the source-to-source spread of the calculated hemispherical emissivity.

\section{Technical Validation}
\label{tech_val}

SpaderOptics is constructed on the basis of refractive indices $n$ and extinction coefficients $k$, obtained from the RefractiveIndex.INFO database \cite{Polyanskiy2024-cv}. RefractiveIndex.INFO database compiles refractive index measurements and estimations reported in the published literature. Since the refractive index and extinction coefficient datasets included in our database originate from numerous studies published over many decades, it is not feasible to independently validate each individual record. Instead, the reliability of the input data is supported by the peer-review process associated with the original literature and by the established expertise of the manufacturers that reported the corresponding optical properties. We further rely on the accuracy of the data extraction, digitization, and conversion procedures, incorporated into the RefractiveIndex.INFO database during its development by Polyanskiy \cite{Polyanskiy2024-cv}.

Accordingly, the technical validation presented here focuses on the aspects that we control, namely:

\begin{itemize}
\item verification of the original data,
\item verification of the resampled data,
\item resampling error analysis,
\item benchmarking against analytical limiting cases, and
\item comparison with experiments for representative cases.
\end{itemize}

\subsection{Verification of the original data}
\label{sec:verification}

Before proceeding to calculations of the optical response quantities, a set of sanity checks was performed on the raw optical datasets imported from the RefractiveIndex.INFO YAML files. The objective of 
these checks was to verify the physical consistency of the input optical constants ($n$ and $k$), and the corresponding wavelength grid provided by each individual literature source. At this stage, no 
interpolation, resampling, or modification of the wavelength grids was introduced by us; each dataset was initially evaluated using the original wavelength sampling 
provided in its corresponding source file. The checks were:

\begin{enumerate}

	\item \textbf{Validity of refractive index and extinction coefficient values.} The imported $n$ and $k$ datasets were checked for non-physical or invalid values: missing values (NaN), infinite values (Inf), negative refractive indices ($n<0$), and negative extinction coefficients ($k<0$) \cite{Reider2016-ic}. Such values can arise from formatting problems, incomplete tables, or numerical artefacts of dispersion formulas. Only a few negative extinction coefficients were found, in the original YAML files from the study of Querry \cite{Querry1985-zb}: 9 for $\mathrm{Al_2O_3}$, 19 for $\mathrm{Cu_2O}$, and 11 for $\mathrm{Fe_2O_3}$. These points were removed; they amount to less than 0.04\% of all extinction coefficient points in the database.
	
	\item \textbf{Monotonicity of the wavelength grid.} The original wavelength grids were checked for a monotonically increasing sequence. Non-monotonic grids can result from inconsistent ordering of the tabulated points within a literature source. Only a few such cases were found, in the original YAML files from the study of Querry \cite{Querry1985-zb}: 1 for $\mathrm{Al_2O_3}$, 1 for $\mathrm{Fe_2O_3}$, and 2 for $\mathrm{Fe_3O_4}$. In these cases, the wavelength points were sorted in ascending order, and the same reordering was applied to the corresponding $n$ and $k$ values, so that each wavelength keeps its own $n$ and $k$.
	
	\item \textbf{Duplicate wavelength entries.} The datasets were also checked for repeated wavelength values with different refractive index values. Such duplicates can occur when a source file contains several measurements or data blocks. Only a few cases were found in the original YAML files: 6 from Yang et al. \cite{Yang2015-gt} for $\mathrm{Ag}$, 3 from Brimhall et al. \cite{Brimhall2009-fj} for $\mathrm{Cu}$, and from Querry \cite{Querry1985-zb}; 2 for $\mathrm{Al_2O_3}$, 1 for $\mathrm{Cu}$, 2 for $\mathrm{Cu_2O}$, and 2 for $\mathrm{Fe_2O_3}$. These points amount to less than 0.02\% of all complex refractive index points in the database. Each group of $N_i$ entries at the same wavelength $\lambda_i$ was replaced by a single point whose optical constants are the arithmetic means of the group:
	\begin{equation}
		n_\mathrm{new}(\lambda_i)=\frac{1}{N_i}\sum_{j=1}^{N_i}n_j(\lambda_i),
		\qquad
		k_\mathrm{new}(\lambda_i)=\frac{1}{N_i}\sum_{j=1}^{N_i}k_j(\lambda_i).
	\end{equation}
	\end{enumerate}
	
\subsection{Verification of the resampled data}
\label{sec:resampled_check}

Following the corrections of the imported optical constants, the optical datasets were combined with the AM0 solar spectrum. Since the wavelength sampling of the literature datasets generally differs from that of the solar spectrum, an additional verification stage was introduced to establish wavelength grid compatibility and ensure that the final optical and thermal-radiative quantities were consistently evaluated on the solar wavelength grid.

Specifically, the following checks were carried out:
\begin{enumerate}
\item \textbf{Wavelength overlap.} The common interval $[\lambda_{s,\min},\lambda_{s,\max}]$ between the wavelength spectrum available by each source and the full AM0 solar spectrum  is defined by Eq.~(\ref{eq:lambda_limits}). A source with $\lambda_{s,\max}\leq\lambda_{s,\min}$ has no overlap with the AM0 spectrum and was excluded. This happened for two sources only: Larruquert et al. \cite{Larruquert1996-po,Larruquert1995-sw} for Al and Brimhall et al. \cite{Brimhall2009-fj} for Cu.

\item \textbf{Solar spectral coverage.} The existence of a wavelength overlap does not necessarily imply that the available optical data cover the full solar wavelength spectrum. To quantify the extent to which each literature source represents the full AM0 solar spectrum, we introduced the solar coverage ratio $\eta_s$, calculated by:
\begin{equation}
\eta_s=\frac{Q_\mathrm{solar}}{Q_\mathrm{AM0}},\qquad
Q_\mathrm{solar}=\int_{\lambda_{s,\min}}^{\lambda_{s,\max}}I_\mathrm{solar}\,d\lambda_s,\qquad
Q_\mathrm{AM0}=\int_{\lambda_{\min}^\mathrm{AM0}}^{\lambda_{\max}^\mathrm{AM0}}I_\mathrm{solar}\,d\lambda
\label{eq:spectral_coverage}
\end{equation}
$Q_{\mathrm{solar}}$ is the solar irradiance contained within the common wavelength interval $[\lambda_{s,\min},\lambda_{s,\max}]$. The limits $\lambda_{s,\min}$ and $\lambda_{s,\max}$ are those estimated by Eq.~(\ref{eq:lambda_limits}). $Q_\mathrm{AM0}$ is the solar irradiance contained within the complete AM0 spectrum $[\lambda_{\min}^{\mathrm{AM0}},\lambda_{\max}^{\mathrm{AM0}}]$. The solar coverage ratio is stored in the database for every source. It is the fraction of the AM0 solar spectrum that lies inside the wavelength range of the source, so it informs the user on how suitable the source is for full-spectrum solar flux calculations. We did not discard sources only because of limited spectral coverage. Their optical data, the calculated optical properties, and the solar-flux quantities are all kept in the database. These datasets remain physically meaningful over the wavelength ranges covered by the respective sources and may provide valuable information for wavelength-specific analyses and comparisons.

\item \textbf{Thermal spectral coverage.} To assess the extent to which the available wavelength range of each source represents the thermal emission of each material at a given temperature, we introduced the thermal coverage ratio. The ratio is defined, as function of temperature, as the fraction of blackbody spectral radiance contained within the wavelength range of the source:
\begin{equation}
\eta_\mathrm{th}(T)=\frac{\displaystyle\int_{\lambda_{s,\min}}^{\lambda_{s,\max}}B_{\lambda,\mathrm{bb}}(\lambda_s,T)\,d\lambda_s}{\displaystyle\int_{0}^{\infty}B_{\lambda,\mathrm{bb}}(\lambda_s,T)\,d\lambda_s}
\label{eq:thermal_coverage_ratio}
\end{equation}
The limits $[\lambda_{s,\min}$ and $\lambda_{s,\max}]$ are the same as for $\eta_s$. The denominator is the full blackbody radiance at all wavelengths, $\sigma_BT^4/\pi$. Thermal coverage ratio is stored for every temperature. It is reported as a diagnostic metric of spectral coverage rather than used as an exclusion criterion. No source or temperature range was removed on the basis of low thermal coverage ratio. Instead, the ratio is provided to allow users to assess the extent to which the available wavelength range represents the thermal emission at the temperatures considered. In particular, lower coverage ratios indicate that a larger fraction of the thermal spectrum lies outside the wavelength range for which optical properties are available by the source. Consequently, emissivity estimates at these temperatures should be used with care.

\item \textbf{Optical and thermal properties verification.} We checked that the resampled $n$ and $k$ values are finite and non-negative. For the calculated quantities, we checked the physical bounds ($0\leq R_{dir},T_{r,dir},A_{dir},\varepsilon_{dir}\leq1$ and $0\leq R_{hemi},T_{r,hemi},A_{hemi},\varepsilon_{hemi}\leq1$), and the energy conservation ($R_{dir}+T_{r,dir}+A_{dir}=1$ and $R_{hemi}+T_{r,hemi}+A_{hemi}=1$). No violations were found.

\item \textbf{Solar flux balance.} For every source, angle, and thickness, we compared the sum of the absorbed, reflected, and transmitted fluxes ($Q_\mathrm{abs,dir}$ $+$ $Q_\mathrm{ref,dir}$ $+$ $Q_\mathrm{tr,dir}$ and $Q_\mathrm{abs,hemi}$ $+$ $Q_\mathrm{ref,hemi}$ $+$ $Q_\mathrm{tr,hemi}$) with the corresponding incident solar flux over the same common wavelength interval. For each source, incidence angle, and thickness, the energy balance residual was calculated as the absolute difference between these quantities. A residual threshold of $tol_Q \leq 10^{-3}\ \mathrm{Wm^{-2}}$ was adopted as the criterion for numerical consistency. This tolerance was selected to account for the numerical precision associated with the wavelength integration and floating-point arithmetic. No violations of this criterion were detected.
\end{enumerate}

\subsection{Resampling error analysis}
\label{sec:Resampling_error}

Resampling is based on interpolations \cite{Good2005-ey}, which introduce errors to
the calculated refractive indices and extinction coefficients compared to the original ones, retrieved from literature. The purpose of
this analysis is therefore to determine whether the adopted solar wavelength grid preserves the original optical-property data with sufficient accuracy for the intended optical and thermal – radiative property calculations. 

To measure the errors, we interpolated the resampled data back to the original wavelength grid. The absolute and relative errors \cite{Taylor2022-zb} are, respectively:

\begin{equation}
\Delta x=x_\mathrm{resampled}-x_\mathrm{original}~~\text{and}~~
\delta_x=\frac{x_\mathrm{resampled}-x_\mathrm{original}}{x_\mathrm{original}}\times100\%
\end{equation}
where $x$ is $n$ or $k$. The mean and standard deviation of both errors for each source are given in a supplementary .txt file, called \fld{Resampling_error_analysis_results.txt}, in the Zenodo repository \fld{https://doi.org/10.5281/zenodo.23643910}~\cite{spaderoptics_data}, allowing users to assess the resampling
fidelity of the individual optical-property datasets.

Overall, the resampling error analysis showed that the maximum relative error stays below 10\%, except for a few sources that include sharp changes of $n$ or $k$ at wavelengths above $20~\mu$m. There the AM0 grid is sparse. Consequently, interpolation over larger wavelength intervals
can introduce larger deviations.

\begin{figure}[h]
    \centering
    \includegraphics[width=1\linewidth]{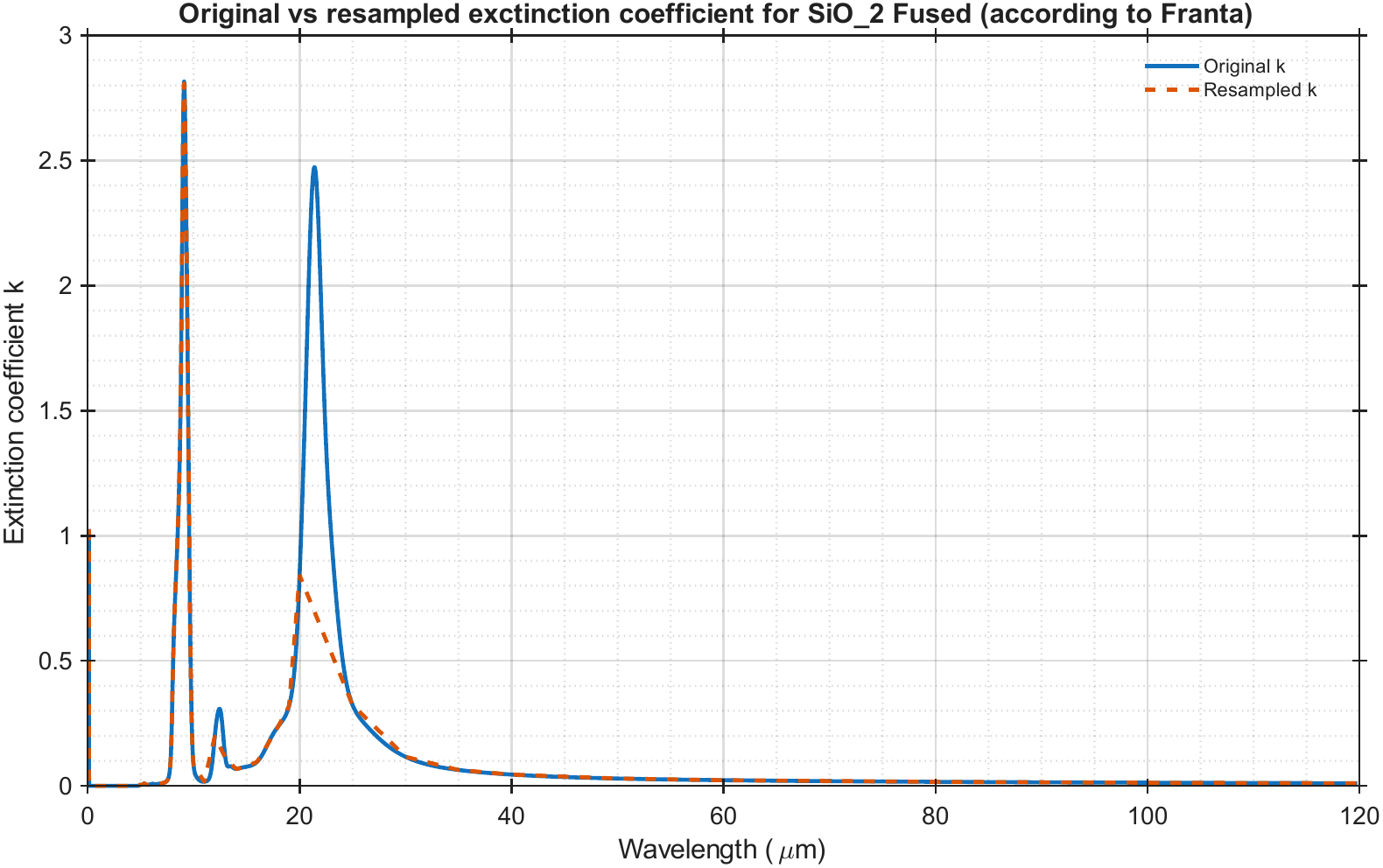}
    \caption{Original (blue, solid) and resampled (orange, dashed) extinction coefficient $k$ for \ce{SiO2} from the study of Franta et al. \cite{Franta2016-jo}.}
    \label{fig:Resampling_error}
\end{figure}

Figure~\ref{fig:Resampling_error} shows an example for \ce{SiO2} from the study of Franta et al. \cite{Franta2016-jo}. The solid blue line is the original $k$. The dashed orange line is $k$ after resampling on the AM0 grid. Below $20~\mu$m, the two lines agree very well. In this range, the AM0 spectrum \cite{ASTM_E490} has very small wavelength steps (0.001 to 1~$\mu$m), so it can follow all the sharp changes of $k$. Above $20~\mu$m, the AM0 grid has fewer points. The step starts at 5~$\mu$m and grows up to 600~$\mu$m. For \ce{SiO2}, the original $k$ (blue line) presents a sharp change between 20 and 25~$\mu$m, but the AM0 grid has only one point at 20~$\mu$m and the next at 25~$\mu$m. The resampled curve (dashed orange line) therefore misses this sharp change. At longer wavelengths, the AM0 grid is even sparser, but $k$ of \ce{SiO2} changes slowly there, so the resampling is accurate again.

Only a few sources show a higher relative error at wavelengths above $20~\mu$m, so we kept this region without adding extra sampling points. This choice balances numerical accuracy against database size: the region has a negligible effect on the solar calculations and a small effect on the thermal emission calculations at high temperatures. Above $20~\mu$m, the AM0 irradiance is more than five orders of magnitude below its peak value \cite{ASTM_E490}, so the solar fluxes are not affected. For thermal emission, we estimated the share of the emitted power that lies above $20~\mu$m using Planck's law \cite{Naterer2021-cn}. The emissive power of a blackbody at temperature $T$ is $E_b = \sigma T^4$. The part of it emitted at wavelengths longer than $\lambda_0$ is
\begin{equation}
	f_{>\lambda_0}(T) = \frac{15}{\pi^4}\int_{0}^{y_0}\frac{x^3}{e^{y}-1}\,\mathrm{d}y,
	\qquad y_0=\frac{hc}{\lambda_0 k_B T}
\end{equation}
where $y=hc/(\lambda k_B T)$ is a dimensionless variable. Setting $\lambda_0=20~\mu$m gives $f_{>20\,\mu\mathrm{m}}\approx1.5\%$ at 1000~K, 0.47\% at 1500~K, and 0.21\% at 2000~K. At these temperatures, which are relevant for concentrated sunlight, this region contributes little to the emitted power. At lower temperatures the share increases, reaching about 9\% at 500~K and 26\% at 300~K, so emissivities computed near room temperature are less accurate.

The sources that include sharp changes at wavelengths above 20$\mu$m and therefore their thermal emission estimates should be used with lower confidence at low temperatures are: Querry \cite{Querry1985-zb} for \ce{Al2O3} and \ce{Fe2O3}, Zhang et al. \cite{Zhang1998-hy} for Kapton, Tsuda et al. \cite{Tsuda2018-gg} for PMMA, Chandler-Horowitz and Amirtharaj \cite{Chandler-Horowitz2005-ei} for Si, Franta et al. \cite{Franta2016-jo} and Herguedas and Carretero \cite{Herguedas2023-vb} for \ce{SiO2} and Synowicki and Tiwald \cite{Synowicki2004-fq} for \ce{ZrO2}.

\subsection{Benchmarking against limiting cases}

Three idealized cases with known solutions were used \cite{Cengel2003-cn}:

\begin{enumerate}
\item \textbf{Ideal lumped blackbody} ($R=0$, $A=1$, $\varepsilon=1$). Under AM0 sunlight, the equilibrium temperature of a lumped body is obtained by balancing the full solar flux with the blackbody thermal emission according to $Q_\mathrm{AM0}=\sigma_BT_\mathrm{eq}^4$, where $Q_\mathrm{AM0}$ is the AM0 irradiance, which is about 1.366~$\mathrm{kW/m^2}$. The analytical result is $T_\mathrm{eq}=(Q_\mathrm{AM0}/\sigma_B)^{1/4}\approx394$~K. The numerical model reproduces this value.
\item \textbf{Ideal mirror} ($R=1$, $A=0$). No solar energy is absorbed, so the temperature stays at its initial value. The numerical model reproduces this ($\Delta T_{\mathrm{eq}}=0$).
\item \textbf{Negligible conduction resistance.} In the limit of negligible conduction resistance, the slab model reduces to the isothermal balance of Section~\ref{sec:lumped}, $T_{\mathrm{f}}=T_{\mathrm{b}}=T_{\mathrm{eq}}$, which the solver returns directly.
\end{enumerate}

\subsection{Comparison with experiments}
\label{sec:experimental_validation}

A comparison with independent experiments tests how well the model, which is used for the database development, reproduces real materials. A full validation of all entries is not feasible, because a meaningful comparison needs measurements with about the same temperature, surface roughness, and oxidation state as the optical data. Such datasets are not systematically
available for all materials and conditions represented in the database. Nevertheless, the database is generated using a consistent physics-based modeling framework across all materials and literature sources, without fine-
tuning or introducing material-specific fitting coefficients. Consequently, representative experimental comparisons can provide an informative assessment
of the predictive capability of the overall modeling approach, while avoiding the implication of comprehensive validation of every individual database entry.

We therefore use two kinds of experimental data. The first is the spectral reflectivity and emissivity of \ce{Au}, \ce{Ag} and \ce{Al} films evaporated in ultrahigh vacuum, measured at room temperature for the wavelength range 0.4--32~$\mu$m by Bennett and Ashley \cite{Bennett1965-aq} (also available in the EKHI database \cite{Gabirondo-Lopez2026-dw}). The second is the hemispherical emissivity measured as a function of temperature by Ramanathan et al. \cite{Ramanathan1977-zb}, which covers about 175--720~K for \ce{Al} and 330--600~K for \ce{Ag}. 

We quantify the agreement with the relative error

\begin{equation}
	\delta_\mathrm{model}=\frac{\varepsilon_\mathrm{calc}-\varepsilon_\mathrm{exp}}{\varepsilon_\mathrm{exp}}\times100\%,
	\label{eq:err}
\end{equation}

where $\varepsilon_\mathrm{calc}$ is the calculated emissivity and $\varepsilon_\mathrm{exp}$ is the measured one. The thresholds $\pm10\%$ and $\pm20\%$ are indicators of agreement, as in earlier studies \cite{ALEXOPOULOU2022877,Alexopoulou2024-if}. They are not general criteria for the quality of a dataset. 

It is noted that the emissivity of good metallic reflectors is small (0.005--0.04), so the relative error is very sensitive to small absolute differences. An absolute difference of 0.001, which is the stated accuracy of the reflectivity measurements of Bennett and Ashley \cite{Bennett1965-aq}, is already a relative error of 10\% at an emissivity of 0.01. The relative errors below should be read with this in mind.

\subsubsection{Comparison with measured spectral data}
\label{sec:bennett}

EKHI \cite{Gabirondo-Lopez2026-dw} is an open database of optical and thermal-radiative properties of solid materials, such as the optical constants and the spectral reflectivity and emissivity, compiled from experimental measurements. In contrast to our database, which is derived from physical models, EKHI is based on measured values. Because it is the closest comparable database, it is a meaningful starting point for the comparison. Specifically, we compare the spectral reflectivity and emissivity of our database with the measurements of Bennett and Ashley \cite{Bennett1965-aq}. As EKHI contains only the reflectivity measurements of Bennett and Ashley for \ce{Ag} and \ce{Au}, we took the emissivities of \ce{Ag} and \ce{Au}, and the reflectivity and emissivity of \ce{Al}, from the original paper.

\begin{figure}[H]
	\centering
	\includegraphics[width=0.95\textwidth]{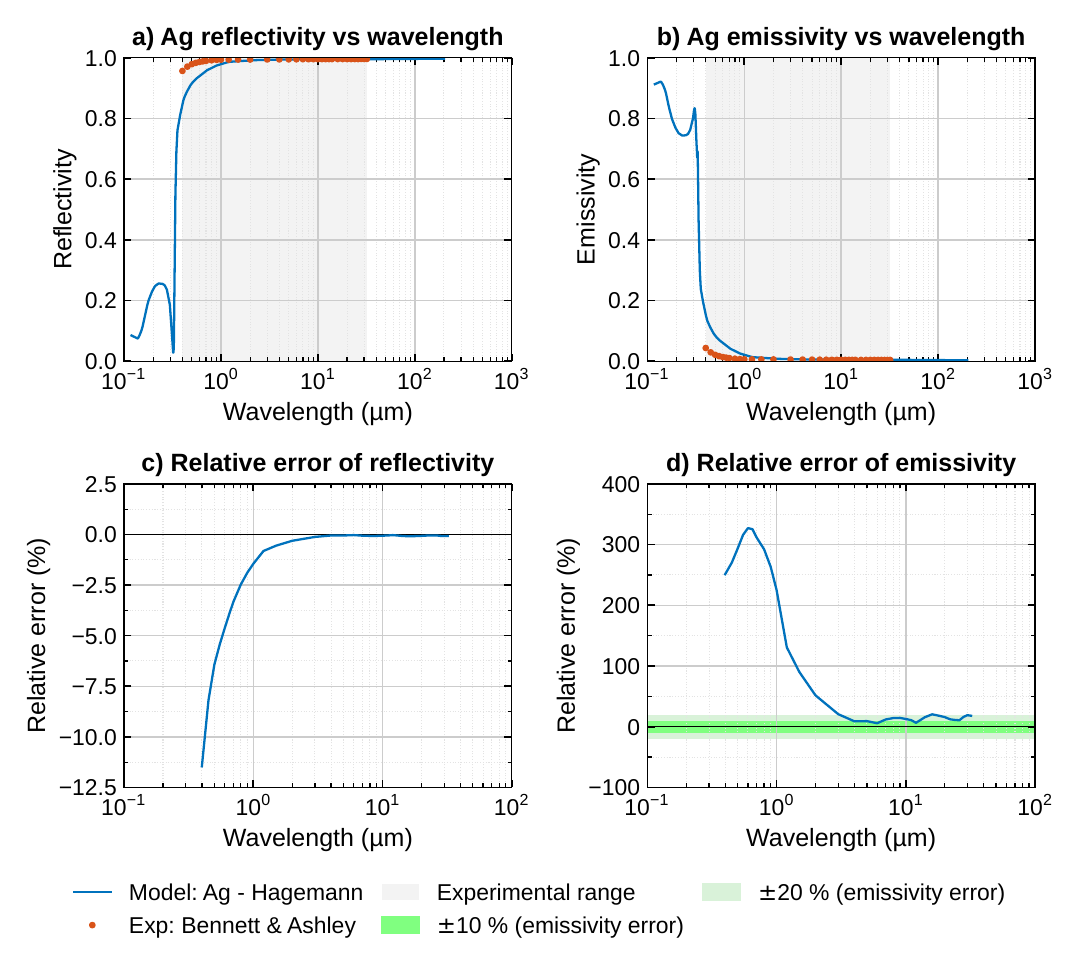}
	\caption{Reflectivity and emissivity of silver from the database (Hagemann) and from the measurements of Bennett and Ashley \cite{Bennett1965-aq} (a, b), and the relative error (c, d) within the common wavelength range of the experiments and the model. The grey area in a, b marks the wavelength range of the measurement; the green areas mark the $\pm10\%$ and $\pm20\%$ relative error thresholds.}
	\label{fig:Bennett_Ag}
\end{figure}

\begin{figure}[t]
	\centering
	\includegraphics[width=0.95\textwidth]{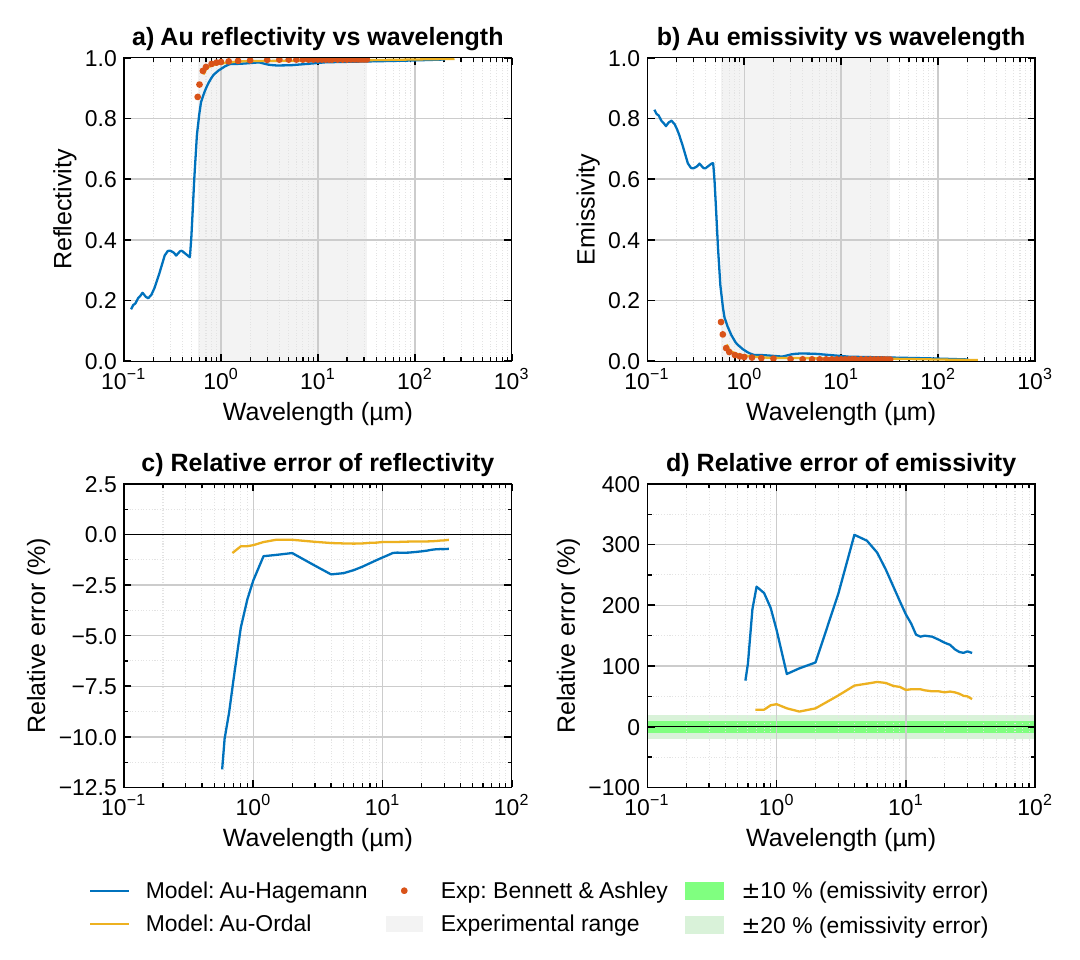}
	\caption{As Figure~\ref{fig:Bennett_Ag}, for gold (Hagemann and Ordal datasets).}
	\label{fig:Bennett_Au}
\end{figure}

\begin{figure}[p!]
	\centering
	\includegraphics[width=0.95\textwidth]{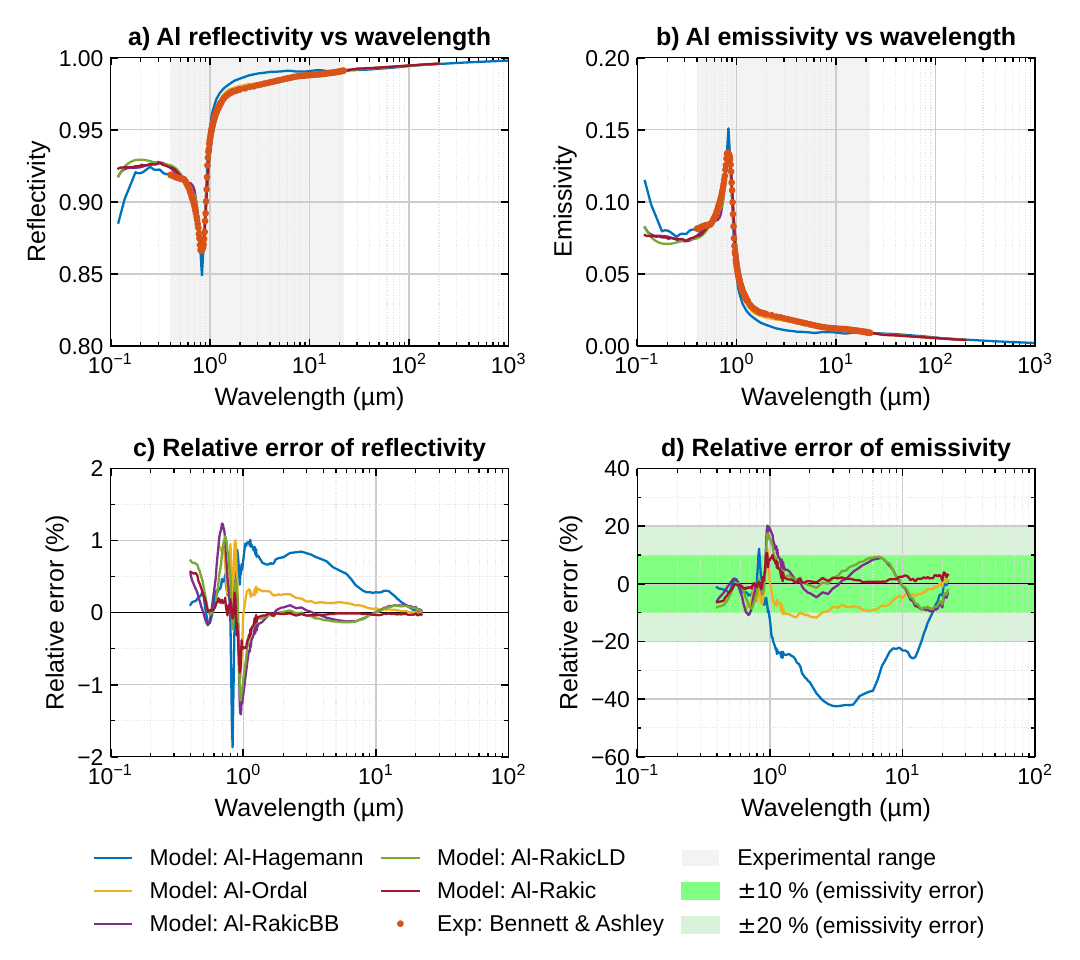}
	\caption{As Figure~\ref{fig:Bennett_Ag}, but for aluminum (five datasets; the measured curve is digitized from the original figure). The database contains eight datasets for aluminum, but only the five that cover the full wavelength range of the experiment, or more, are shown: one from Hagemann et al.~\cite{Hagemann1975-sg}, one from Ordal et al.~\cite{Ordal1988-me}, and three from Rakic et al.~\cite{Rakic1998-xo}. The remaining datasets, those of Cheng et al.~\cite{Cheng2016-cr}, Mathewson and Myers~\cite{Mathewson1971-fv} and McPeak et al.~\cite{McPeak2015-or}, do not cover the full wavelength range of the experiment and are therefore excluded from this comparison.}
	\label{fig:Bennett_Al}
\end{figure}

They measured the near-normal (5$^\mathrm{o}$ incidence angle) reflectivity $R_{dir}$ of opaque silver and gold films evaporated in ultrahigh vacuum (UHV), with an accuracy of about $\pm0.1\%$. These reflectivities are also included in the EKHI database \cite{Gabirondo-Lopez2026-dw}. They give the emissivity as $1-R_{\mathrm{dir}}$. The paper tabulates silver (for the wavelength range 0.4--32~$\mu$m) and gold (for the wavelength range 0.575--32~$\mu$m). For aluminum, only a figure is available, and we digitized the curve of the aluminum film (for the wavelength range 0.4--22~$\mu$m). For these comparisons, we use only those datasets from our database that cover the full wavelength range of the experimental datasets. Figures~\ref{fig:Bennett_Ag}--\ref{fig:Bennett_Al} show the results of the comparisons.

For silver, the Hagemann dataset reproduces the measured reflectivity within 0.1\% beyond 3~$\mu$m, which is within the accuracy of the measurement, and within 1.4\% at 1--3~$\mu$m. The deviation grows towards the short wavelengths and reaches $-11\%$ at 0.4~$\mu$m, near the interband absorption edge. The emissivity is $+6$ to $+20\%$ above the measurement beyond 3~$\mu$m, and much higher at shorter wavelengths (more than 200\% below 1~$\mu$m), because the emissivity there is small and the reflectivity error is relatively large.

For gold, the datasets differ more. The reflectivity of the Hagemann dataset is 0.7--2.3\% below the measurement at 1--32~$\mu$m, which gives an emissivity two to four times that of the measurement ($+80$ to $+320\%$). The Ordal dataset is closer: the reflectivity is 0.3--0.9\% below the measurement, and the emissivity is $+25$ to $+74\%$ above it. The two gold datasets therefore differ in emissivity by a factor of about two in the thermal infrared.

For aluminum, all reflectivity datasets have relative errors that lie between -2\% and 1.5\%. For example, the Rakic datasets reproduce the reflectivity within about 1\% at 0.4--3~$\mu$m and within 0.1\% beyond 3~$\mu$m. The emissivity of the Rakic dataset is between $-6\%$ and $+11\%$ of the measurement over the whole range. The Ordal dataset is within $\pm12\%$. The Hagemann dataset is $-12$ to $-42\%$ below the measurement at 1--10~$\mu$m, which agrees with its large errors against the temperature measurements.

To separate the deviation of a single dataset from the effect of the choice of dataset, we compare the relative error with respect to the measurement with the spread between the datasets of the same material, defined as the difference between the highest and the lowest emissivity divided by the measured emissivity at each wavelength. The datasets describe materials in different conditions, so a larger deviation does not mean that a dataset is of lower quality, but that its conditions differ more from those of the measured films. For \ce{Al}, the datasets closest to the measurement (Rakic and Ordal), deviate from it by a median of 2--7\%, whereas the spread between the five datasets is 28\% (median; 4--48\% depending on wavelength), mostly because of the Hagemann dataset (without it, 12\%). The choice of the dataset therefore has a larger effect than the deviation of the dataset closest to the measurement. For \ce{Au}, all datasets overestimate the emissivity (median $+57\%$ for Ordal and $+150\%$ for Hagemann), and the spread between the two datasets (median 90\%, up to 250\%) is larger than the deviation of the dataset closer to the measurement. For \ce{Ag}, only one dataset is compared with the spectral measurement, so no spread can be given; its median deviation is 13\% beyond 3~$\mu$m.

\subsubsection{Aluminum: temperature dependence}

\begin{figure}[h]
	\centering
	\includegraphics[width=\linewidth]{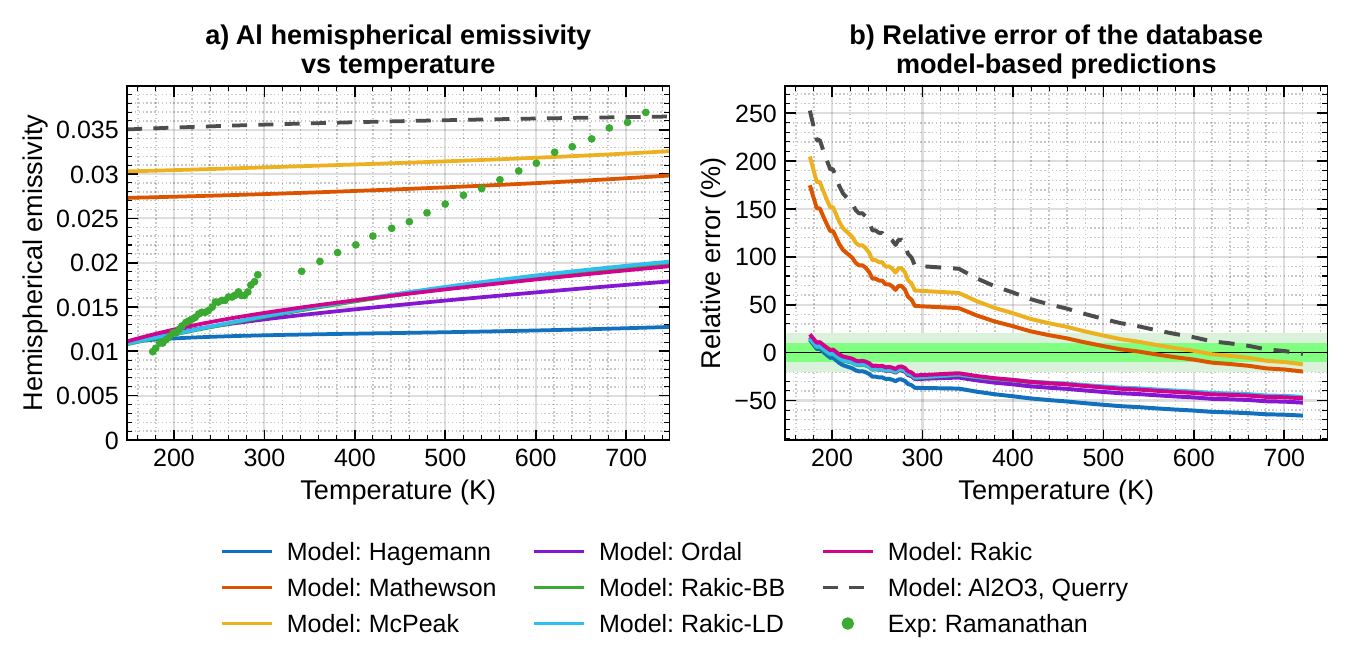}
	\caption{(a) Hemispherical emissivity of aluminum: calculated from different datasets of the database and measured by Ramanathan et al. \cite{Ramanathan1977-zb}. The \ce{Al2O3} curve (Querry \cite{Querry1985-zb}) is a reference for a fully oxidized surface. (b) Relative error between the calculated and the measured emissivity (Eq.~\ref{eq:err}). The intense and pale green areas mark the $\pm10\%$ and $\pm20\%$ relative error thresholds.}
	\label{fig:Al}
\end{figure}

Figure~\ref{fig:Al}a compares hemispherical emissivity of seven aluminum datasets with the measurements of Ramanathan et al. \cite{Ramanathan1977-zb}, who used electropolished aluminum of 99.999\% purity. The oxidation state of the samples is not documented by Ramanathan et al. \cite{Ramanathan1977-zb}. The aluminum datasets have no temperature-dependent optical constants, so the temperature dependence of the calculated emissivity comes only from the Planck weighting (Section~\ref{sec:emissivity}). The database contains eight sources for aluminum. In this comparison, we include seven of them: one dataset each from Hagemann et al.~\cite{Hagemann1975-sg}, Ordal et al.~\cite{Ordal1988-me}, Mathewson and Myers~\cite{Mathewson1971-fv}, and McPeak et al.~\cite{McPeak2015-or}, and three datasets from Rakic et al.~\cite{Rakic1998-xo}. The dataset of Cheng et al.~\cite{Cheng2016-cr} is the only one excluded, because its experimental conditions (epitaxially grown, atomically smooth aluminum layers) differ significantly from those of the experiment of Ramanathan et al.~\cite{Ramanathan1977-zb}. As a reference for a fully oxidized surface, we added \ce{Al2O3} from the dataset of Querry \cite{Querry1985-zb}.

Each dataset reproduces the measurements within $\pm20\%$ only over a limited temperature range, and every dataset may deviate by more than 20\% outside it (Figure~\ref{fig:Al}b). The Hagemann \cite{Hagemann1975-sg}, Ordal \cite{Ordal1988-me}, and Rakic \cite{Rakic1995-na,Rakic1998-xo} datasets are within about $\pm20\%$ only at low temperatures. The deviation then grows and reaches about $-45$ to $-65\%$ at 700~K. In contrast, the Mathewson and Myers \cite{Mathewson1971-fv} and McPeak \cite{McPeak2015-or} datasets and the \ce{Al2O3} curve by Querry \cite{Querry1985-zb} overestimate the emissivity by $+175$ to $+250\%$ at low temperatures, but come within $\pm20\%$ at about 450~K (Mathewson and Myers), 490~K (McPeak), and 600~K (\ce{Al2O3}). No dataset stays within $\pm20\%$ over more than about half of the measured range. Statements that a dataset agrees with the experiment are therefore valid only for the appropriate temperature range.

The large relative errors at low temperature arise partly because the measured emissivity is smallest there (0.010 at 175~K), so the same absolute deviation produces a large relative error. For the Mathewson and Myers dataset, for example, the absolute deviation is $+0.017$ at 175~K and $-0.007$ at 720~K, which corresponds to a relative error of about $+175\%$ at 175~K but only about $-20\%$ at 720~K.

The datasets also describe different surface states. The first group (Hagemann, Ordal, Rakic) describes pure aluminum, or films measured before significant oxidation. The second group (Mathewson and Myers, McPeak) comes from experimentally measured films and therefore includes some oxide and surface effects. Aluminum oxidizes easily, even in high vacuum during film deposition \cite{McPeak2015-or}, so nominally pure samples can carry an oxide layer, which grows with the duration of the experiment and thus with temperature. The \ce{Al2O3} curve shows the limit of a fully oxidized surface. Although Ramanathan et al.\ \cite{Ramanathan1977-zb} do not state the oxidation state of their samples, the observed increase of the emissivity with temperature is consistent with the expected influence of oxidation and with the high reactivity of aluminum reported in the literature \cite{McPeak2015-or}. The datasets of our database are consistent with this explanation within their respective suitable temperature ranges.

The choice of dataset therefore has a large effect on the result, and a good choice requires prior knowledge of the surface state. Users should compare sources and use their provenance to decide which source best suits the needs of their specific application.

\subsubsection{Silver: temperature dependence}

Optical properties of silver depend strongly on temperature, making it particularly suitable for demonstrating the importance of retaining
temperature-dependent optical-property data in our database. Temperature enters the emissivity in two ways. First, the Planck weighting changes with temperature (Eqs.~\ref{eq:planck}--\ref{eq:Qemit}). This effect applies to all sources, whether or not their $n$ and $k$ depend on temperature. Second, for sources with temperature-dependent $n$ and $k$, the spectral absorptivity, and therefore the emissivity, also changes with temperature. A dataset given at a single temperature does not include this second effect.

\begin{figure}[!p]
	\centering
	\includegraphics[width=\linewidth]{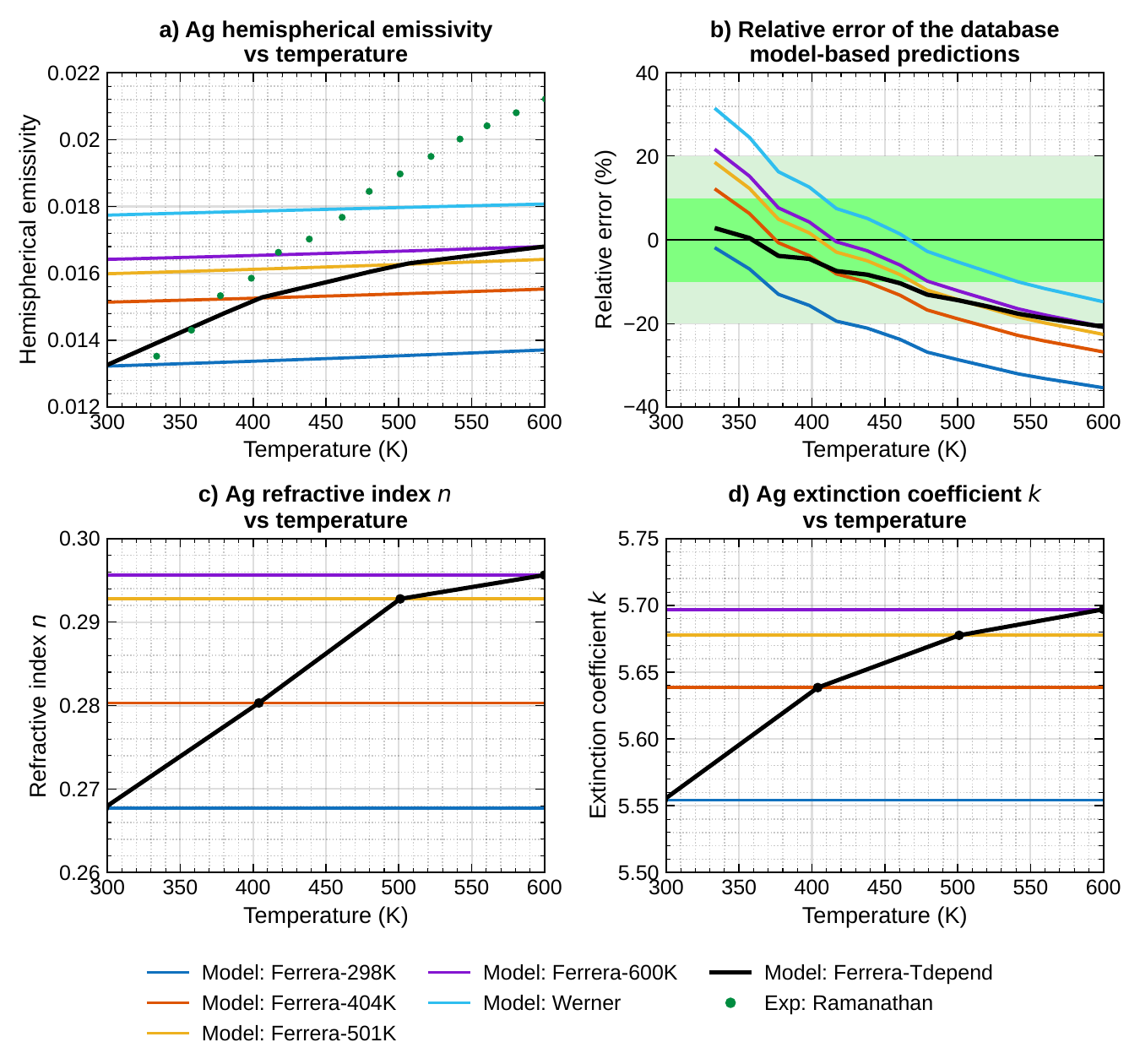}
	\caption{(a) Hemispherical emissivity of silver: calculated from the temperature-specific datasets of Ferrera et al. \cite{Ferrera2019-ed}, from the combined ``Ferrera-Tdepend'' entry, and from the dataset of Werner et al. \cite{Werner2009-me}, and measured by Ramanathan et al. \cite{Ramanathan1977-zb}. (b) Relative error between the calculated and the measured emissivity. The intense and pale green areas mark the $\pm10\%$ and $\pm20\%$ relative error thresholds. (c, d) Refractive index $n$ and extinction coefficient $k$ of silver from the Ferrera et al. \cite{Ferrera2019-ed} datasets, integrated over the measured wavelength range. The horizontal lines show the values of the individual temperature-specific datasets, assumed constant with temperature, and the black line connects the values of these datasets at their own temperatures, i.e., the temperature-dependent case (``Ferrera-Tdepend'').}
	\label{fig:Ag}
\end{figure}

We compare the dataset of Ferrera et al. \cite{Ferrera2019-ed} with the measurements of Ramanathan et al. \cite{Ramanathan1977-zb}. Ferrera et al. \cite{Ferrera2019-ed} dataset is given for four temperatures (298~K, 404~K, 501~K, 600~K) in the RefractiveIndex.INFO database \cite{Polyanskiy2024-cv}. Our database has one entry per temperature and the combined ``Ferrera-Tdepend'' entry. Figure~\ref{fig:Ag}a shows the emissivity, and Figure~\ref{fig:Ag}b the relative error.

For completeness, Figure~\ref{fig:Ag}c,d shows the temperature dependence of the optical constants of silver underlying the Ferrera et al. \cite{Ferrera2019-ed} predictions. For each dataset, $n$ and $k$ were integrated over the measured wavelength range. Between 298 and 600~K, $n$ increases from 0.268 to 0.296 (about 10\%) and $k$ from 5.55 to 5.70 (about 3\%). The horizontal lines show the values of the individual temperature-specific datasets, assumed constant with temperature, whereas the black line, which connects the values of these datasets at their own temperatures, represents the temperature-dependent case (``Ferrera-Tdepend''). This increase of $n$ and $k$ with temperature is what raises the Ferrera-Tdepend emissivity relative to any single-temperature dataset (Figure~\ref{fig:Ag}a), and it accounts for the improved agreement with the measured temperature trend.

The single-temperature curves of Ferrera et al.\ \cite{Ferrera2019-ed} are calculated with constant $n$ and $k$. They agree well close to their own temperature, but the error grows away from it and exceeds 30\% in the worst case. The combined entry ``Ferrera-Tdepend'' describes the data better over a wider temperature range, since it reproduces the rising trend of the measured emissivity, although it still does not follow the measurement over the full range. Its error changes from about $+3\%$ at 335~K to about $-21\%$ at 600~K, staying within $\pm10\%$ up to about 450~K and within $\pm20\%$ up to about 590~K. The measured emissivity increases by 57\% between 333 and 600~K, whereas the calculated one increases by only about 26\%.

This divergence comes from the different conditions of the two experiments. Ferrera et al.\ \cite{Ferrera2019-ed} deposited an optically thick Ag film by molecular beam epitaxy and measured it in ultrahigh vacuum (UHV, below $10^{-9}$~Torr), where both surface roughness and contamination are small \cite{Stahrenberg2001-nk}. Ramanathan et al. \cite{Ramanathan1977-zb}, in contrast, measured a mechanically polished bulk Ag sample, which has a considerably rougher surface, in a non-ultrahigh vacuum of $10^{-6}$~Torr. In such a vacuum, contamination becomes a major factor, according to Werner et al.\ \cite{Werner2009-me}. Ramanathan et al.\cite{Ramanathan1977-zb} do not report the roughness or the contamination of their sample, but surfaces prepared and measured under these conditions are known to be rough and contaminated, according to the literature \cite{Stahrenberg2001-nk,Bennett1965-aq}.

For silver we refer to contamination rather than oxidation, unlike to aluminum. Silver oxide (\ce{Ag2O}) is thermodynamically unstable above about 460~K (190~$^\circ$C) in air at atmospheric pressure \cite{CRC}, and the measurements were done at about $10^{-6}$~Torr, where it decomposes at an even lower temperature. An oxide layer is therefore not expected to grow on silver in the temperature range where the deviation increases, whereas aluminum has a stable native oxide. Other surface changes are more probable. Silver sulfide, for example, forms from traces of sulfur compounds and does not decompose on heating.

To examine the effect of roughness, we also compared the measurement of Ramanathan et al.\ \cite{Ramanathan1977-zb} with the results that our database gives for the data of Werner et al.\ \cite{Werner2009-me}. They used bulk Ag and, in addition to polishing, applied 3~keV Ar$^+$ ion bombardment to remove contaminants before the measurement. Since the experiment was carried out in UHV, no significant contamination builds up during it, as reported by Bennett and Ashley \cite{Bennett1965-aq}, so the effect of roughness can be examined almost independently. The Werner et al. \cite{Werner2009-me} dataset goes from $+32\%$ at 335~K to $-15\%$ at 600~K. The larger deviation at low temperature can be attributed to the ion bombardment, which, according to the literature \cite{Stahrenberg2001-nk}, adds to the roughness caused by polishing alone. As the temperature increases, this additional roughness is smoothed out because it is very shallow. This can explain why the Werner et al. \cite{Werner2009-me} dataset is closest to the measurement from about 450~K onwards, where both surfaces are rough only because of their initial polishing.

Even with the roughness effect included as in the Werner et al. \cite{Werner2009-me} data, the emissivity measured by Ramanathan et al.\cite{Ramanathan1977-zb} keeps rising with temperature. Roughness can no longer explain this rise, but contamination, which was not removed in their experiment, can. Unlike the sources in our database, which were measured in UHV, Ramanathan et al.\cite{Ramanathan1977-zb} worked in a non-UHV environment, where contamination grows with the duration of the experiment and therefore with temperature. To the best of our knowledge, no other experiment has measured the emissivity of silver with contamination, so we could not compare directly with Ramanathan et al.\cite{Ramanathan1977-zb} at their highest temperatures. Bennett and Ashley \cite{Bennett1965-aq} do report, however, that in non-UHV measurements contamination becomes the most significant factor, more significant than roughness.

\subsubsection{Variability between sources}
\label{sec:spread}

The comparisons above show that there is significant source-to-source variability: for the same material, the original $n$ and $k$ data and therefore all the calculated quantities can differ substantially between literature sources. To quantify this variability for every material, we included in the database the structure \fld{source_spread} (Section~\ref{sec:datarecord}), which is computed from the calculated hemispherical emissivity of the sources of each material.

We use the emissivity for three reasons. First, the hemispherical emissivity is stored for every source with the same dimensions, temperature $T \times$ thickness $d$, and on the same temperature and thickness grids, so the sources can be compared point by point without any interpolation. Other quantities do not have this property: the spectral properties depend on the wavelength range covered by each source, and the temperature-dependent (``-Tdepend'') entries carry an additional temperature dimension. Second, the emissivity is a representative quantity of the whole calculation chain. It results from the refractive index and extinction coefficient of the source, through the calculated spectral absorptivity and its Planck weighting (Section~\ref{sec:emissivity}), and it therefore also reflects differences in the wavelength coverage of the sources. Third, the emissivity determines the emitted flux and thus enters the energy balance that gives the equilibrium and slab temperatures. Variability in the emissivity is therefore directly relevant to the temperatures calculated from each source.

For a given material with $N$ sources, let $\varepsilon_s(T,d)$ denote the calculated hemispherical emissivity of source $s=1,\dots,N$ at temperature $T$ and thickness $d$. For each combination of $T$ and $d$, we compare the $N$ values $\varepsilon_1(T,d),$ $\dots,$ $\varepsilon_N(T,d)$ of the sources of that material and calculate the following metrics. Each metric is therefore stored as an array over temperature and thickness, with the same dimensions as the emissivity itself:

\begin{itemize}
	\item The minimum $\varepsilon_\mathrm{min}(T,d)=\min[\varepsilon_s(T,d)]$ and the maximum $\varepsilon_\mathrm{max}(T,d)=\max[\varepsilon_s(T,d)]$ (fields \fld{min} and \fld{max}). They are the lowest and the highest emissivity that any source of the material predicts at that $T$ and $d$, and together they form the envelope of the results.
	\item The mean $\bar{\varepsilon}(T,d)=\frac{1}{N}\sum_{s=1}^{N}\varepsilon_s(T,d)$ (field \fld{mean}). It is the average emissivity of all sources and it is the reference value for the standard deviation in Eq.~\eqref{eq:std}. Unlike the median, it is affected by extreme values.
	\item The median $\varepsilon_\mathrm{med}(T,d)$ (field \fld{median}). The $N$ emissivities $\varepsilon_s(T,d)$ are sorted in ascending order, $\varepsilon_{(1)}\le\varepsilon_{(2)}\le\dots\le\varepsilon_{(N)}$, and the median is the middle value:
	\begin{equation}
		\varepsilon_\mathrm{med}=
		\begin{cases}
			\varepsilon_{\left(\frac{N+1}{2}\right)}, & N \text{ odd},\\[6pt]
			\dfrac{1}{2}\left(\varepsilon_{\left(\frac{N}{2}\right)}+\varepsilon_{\left(\frac{N}{2}+1\right)}\right), & N \text{ even}.
		\end{cases}
		\label{eq:median}
	\end{equation}
	It gives half of the sources above it and half below it, and is the reference value for the median absolute deviation (MAD) in Eq.~\eqref{eq:mad}. Unlike the mean, it is not affected by the size of an extreme value.
	\item The standard deviation $\sigma_\varepsilon$ (field \fld{std})
	\begin{equation}
		\sigma_\varepsilon=\sqrt{\frac{1}{N-1}\sum_{s=1}^{N}\left(\varepsilon_s-\bar{\varepsilon}\right)^2},
		\label{eq:std}
	\end{equation}
	It shows the scatter of all sources around their mean, not only of the two extreme ones.
	\item The median absolute deviation $\mathrm{MAD}_\varepsilon$ (field \fld{mad}) 
	\begin{equation}
		\mathrm{MAD}_\varepsilon=\mathrm{median}\left|\varepsilon_s-\varepsilon_\mathrm{med}\right|
		\label{eq:mad}
	\end{equation}
	It is the median of the absolute distances of the sources from the median. It shows the typical deviation of a source from the typical value, and, like the median, it is not affected by single extreme sources.
	\item The absolute range $\Delta\varepsilon$ (field \fld{range_abs})
	\begin{equation}
		\Delta\varepsilon=\varepsilon_\mathrm{max}-\varepsilon_\mathrm{min}
		\label{eq:range_abs}
	\end{equation}
	It is the width of the envelope defined between $\varepsilon_{\mathrm{min}}$ and $\varepsilon_{\mathrm{max}}$.
	\item The relative range $\Delta\varepsilon_\mathrm{rel}$ (field \fld{range_rel_pct})
	\begin{equation}
		\Delta\varepsilon_\mathrm{rel}=\frac{\varepsilon_\mathrm{max}-\varepsilon_\mathrm{min}}{\varepsilon_\mathrm{med}}\times100\%
		\label{eq:range_rel}
	\end{equation}
	It is the absolute range divided by the median, expressed in percent. The same absolute range has a very different importance depending on the emissivity: for example, a range of $\Delta\varepsilon=0.02$ corresponds to $\Delta\varepsilon_\mathrm{rel}=200\%$ for a median of $0.01$, but only to $5\%$ for a median of $0.4$. In cases where emissivity changes strongly with temperature and thickness, the absolute range alone cannot show where the choice of source matters. The relative range is therefore stored in the database, so that users can identify directly the conditions under which the sources agree (small $\Delta\varepsilon_\mathrm{rel}$) and those under which the choice of the source changes the result significantly (large $\Delta\varepsilon_\mathrm{rel}$).
\end{itemize}

The metrics are calculated for all sources of a material, except the temperature -- dependent entries, because these are combinations of the temperature -- specific sources of the same author and would count one author more than once. A spread is stored only for materials with at least two sources.

Finally, these metrics describe the differences between the sources of the database, which represent materials in different conditions, and not a statistical uncertainty of the emissivity. A large spread does not mean that a source is wrong, but that the choice of source matters for the result.

\section{Usage Notes}

\subsection{Database size and memory footprint}

The file \fld{SpaderOptics.mat} has the MATLAB v7.3 format, which is based on HDF5. It has a size of 1.37~GB on disk and requires 2.81~GB of memory when it is fully loaded into MATLAB. The memory is dominated by the directional reflectivity, transmissivity and absorptivity (\fld{calculated_R_directional}, \fld{calculated_T_directional}, \fld{calculated_A_directional}), which take about 0.87~GB each, i.e., about 92\% of the total. The largest single array, with dimensions $\lambda_s \times \theta_i \times d \times T$, belongs to a temperature-dependent entry of Si (\fld{Franta-Tdepend} \cite{Franta2017-pm}) and has a size of 103~MB. The memory required by one material ranges from about 1~MB to about 1~GB (Si, with 61 sources).

\subsection{Accessing the database}

The database is deposited as a single MATLAB file, \fld{SpaderOptics.mat}, in the -v7.3 format. This format is based on the open HDF5 standard, so the file can be read with MATLAB and also with open-source tools that do not require a MATLAB license. Below we describe how to access the whole database in MATLAB and in Python. Since the complete file has a size of 1.37~GB on disk and requires about 2.81~GB of memory when fully loaded, we also describe how to access a single material, without needing to load the full database. For both MATLAB and Python, a README file is provided with the database. It explains step by step how to load the database and retrieve data, lists the field names, units and dimension order of every array, and gives examples for accessing each type of data, including the \fld{source_spread} structure.

\subsubsection{Accessing through MATLAB}

In MATLAB, the whole database is loaded with the command \fld{load(`SpaderOptics.mat')}, which creates the structure array \fld{SpaderOptics}. Because of the HDF5-based format, parts of the file can also be read without loading the rest, using the command \fld{matfile}. For example,
\begin{center}
	\fld{m} \fld{=} \fld{matfile(`SpaderOptics.mat');}
	
	\fld{material} \fld{=} \fld{m.SpaderOptics(1,i);}
\end{center}
loads only the material with index \texttt{i}, including all its sources. Users who do not need the full database are advised to use this approach. The MATLAB README describes how to find the index of a material and how to select a source, a quantity, a temperature and a thickness.

\subsubsection{Accessing through Python}

To make the database accessible with open-source tools, in line with the FAIR principles, we provide a Python script, \fld{load_spader.py}, which uses the open-source libraries \fld{numpy} and \fld{h5py}. The script loads all materials and sources into nested Python dictionaries, \fld{db[material][author][field]}, and the source spread statistics into a second dictionary, the \fld{spread[material][field]}.

Users with limited memory can read a single material instead of the full database. Since the file is in HDF5 format, \fld{h5py} opens it without loading its content, and only the data of the requested material are then read into memory. The Python README provides a ready-to-use function for this purpose, so that, for example, \fld{load_material(`SpaderOptics.mat', 'Al')} returns the data of all aluminum sources. Together with loading guidance, the Python README describes the field names, units and dimension order of all arrays, as in Section~\ref{sec:datarecord}, and gives examples on how to retrieve each type of data, such as the hemispherical emissivity at a given temperature and thickness, or the spectral reflectivity at a given angle.

\subsection{The SpaderOptics App}

\begin{figure}[p!]
	\centering
	\includegraphics[width=\textwidth]{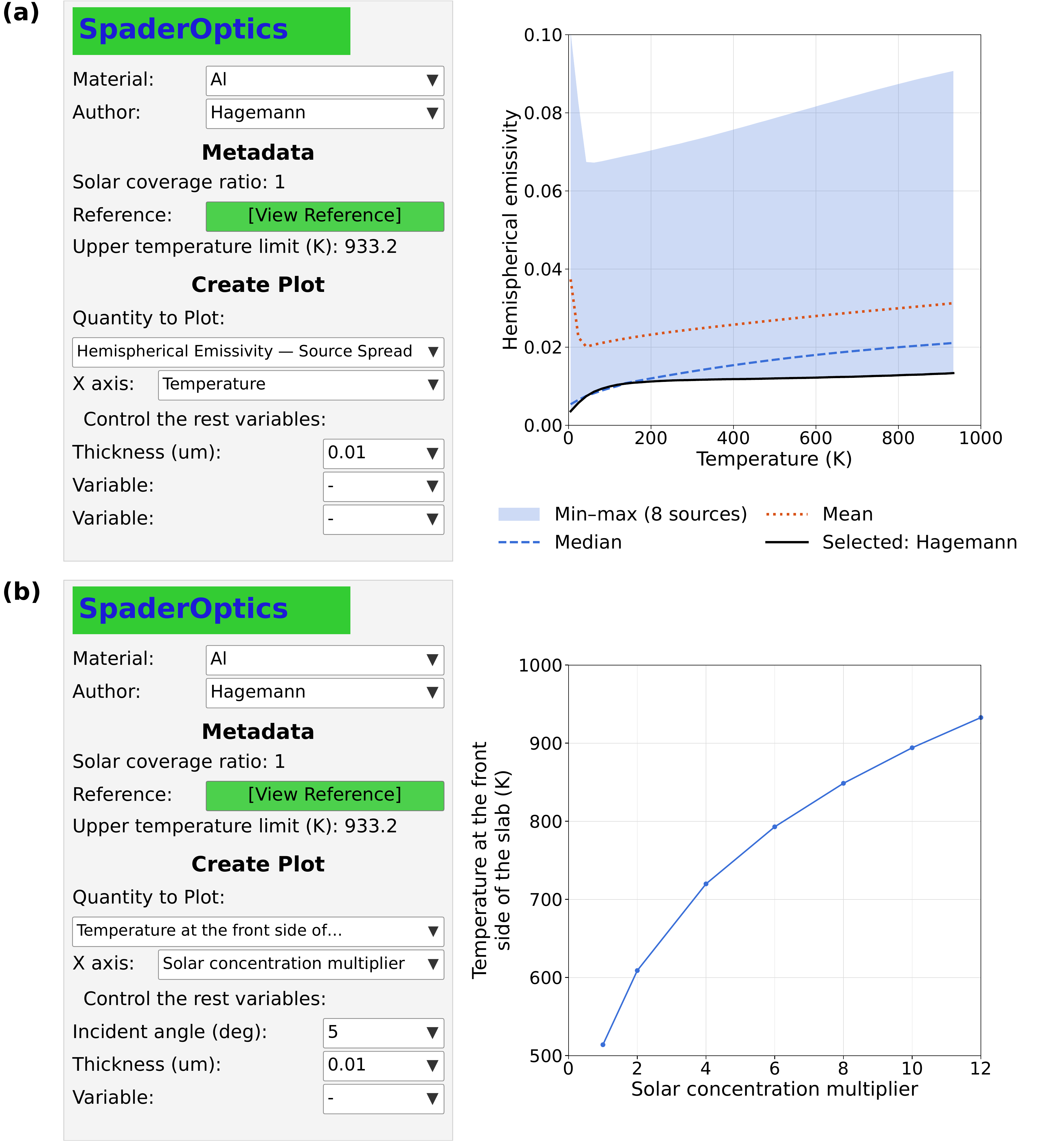}
	\caption{An example case usage of the SpaderOptics App for aluminum (according to the Hagemann source \cite{Hagemann1975-sg}), for thickness $0.01~\mu$m.
		(a)~The interface with the quantity ``Hemispherical Emissivity --- Source Spread''
		and ``Temperature'' on the X axis. The plot shows the min -- max envelope of the eight
		sources (blue, shaded) available in the database, their median (blue, dashed) and mean (orange, dotted), and the emissivity of the
		selected source (black, solid).
		(b)~The interface with the quantity ``Temperature at the front side of the slab''
		and the ``Solar concentration multiplier'' on the X axis, at an incident angle of
		$5^\circ$.}
	\label{fig:app}
\end{figure}

The SpaderOptics App is a graphical interface to the database (Figure~\ref{fig:app}). At the top of the interface, the user selects the \fld{Material} and the literature source (\fld{Author}) from the two drop-down lists. The author list is updated automatically for each material. For the selected material and author, the \fld{Metadata} section shows the solar coverage ratio, the upper temperature limit (in K) and a \fld{[View} \fld{Reference]} button, which opens a window with the literature references of the refractive-index and thermal conductivity data.

In the \fld{Create} \fld{Plot} section, the user first selects the \fld{Quantity} \fld{to} \fld{Plot} from a drop-down list. The available quantities are the refractive index $n$ and the extinction coefficient $k$; the thermal conductivity; the directional and hemispherical reflectivity, transmissivity and absorptivity; the directional and hemispherical emissivity; the directional and hemispherical reflected, transmitted and absorbed solar fluxes and the hemispherical emitted flux; the lumped-body equilibrium temperature; the conductive flux; the temperatures at the front, the rear and the middle of the slab model (Section~\ref{sec:slab}), each for directional or hemispherical absorptivity combined with hemispherical emissivity; the thermal coverage ratio; and the spread of the hemispherical emissivity between the sources of the selected material (Section~\ref{sec:spread}), as the absolute spread and as the relative spread. Next, the user selects the variable of the horizontal axis (\fld{X} \fld{axis}), for example the wavelength, the incident or emission angle, the thickness, the temperature or the solar concentration multiplier. The available choices depend on the quantity to plot. Finally, up to three further drop-down lists, under section \fld{Control} \fld{the} \fld{rest} \fld{variables}, set the values of the remaining parameters, such as the incident angle, the thickness, the temperature or the wavelength that was not selected as X axis. The available options depend on the selected quantity to plot and X axis. The plot, at the right side of the interface is updated every time a selection changes.

Figure~\ref{fig:app} shows two examples from using the SpaderOptics App for aluminum (according to the Hagemann source \cite{Hagemann1975-sg}). In Figure~\ref{fig:app}a, the hemispherical emissivity of the selected source is shown as a function of temperature, together with the minimum--maximum range, the median and the mean of all aluminum sources (Section~\ref{sec:spread}). In Figure~\ref{fig:app}b, the temperature at the front side of the slab, calculated with the directional absorptivity and the hemispherical emissivity, is shown as a function of the solar concentration multiplier, for an incident angle of 5$^\circ$ and a thickness of 0.01~$\mu$m. The curves end at the upper temperature limit of the material (933.2~K for aluminum).

\subsection{Limitations and recommended use}
\label{sec:limitations}

The database takes complex refractive indices from the literature as input and, on this basis, calculates optical, thermal and radiative quantities of the materials using physical models. The range of validity of the input data and of the models therefore defines the range of validity of the database and its limitations. The following limits apply.

\begin{itemize}
	\item \textbf{Solid state only.} Both the literature data and the models used to calculate the optical properties describe the solid state. The database is therefore valid up to the melting point and does not extend to the liquid or gaseous states, since the optical properties of a liquid cannot be derived from solid-state data. For seven materials, either there is no conventional melting point or another process occurs at a lower temperature, above which the literature data or the models are no longer valid. For these materials, the upper temperature included in the database is correspondingly lower; they are listed in Table~\ref{tab:upperlimit}. To inform the user, the upper temperature limit $T_\mathrm{up}$ is stored in the database for every material (field \fld{upper_temperature_limit_K}) and is shown in the App. No property is calculated above $T_\mathrm{up}$.
	
	\item \textbf{Surface roughness and oxidation.} The models used to develop the database do not account for surface roughness or oxidation, nor for their evolution with time or temperature. These effects are included only if they are already contained in the refractive indices reported in the literature. Moreover, each literature source corresponds to a specific roughness or oxidation state and does not describe how that state evolves. The only way to represent increasing roughness or oxidation, as we do in the technical validation for \ce{Al} and \ce{Ag} (Section~\ref{sec:experimental_validation}), is to compare sources from different authors whose samples were in different roughness or oxidation states. When selecting a source, users should therefore consult its provenance and consider the conditions under which it was measured. The further the intended application is from those conditions, the lower the confidence in the database results  should be.
	
	\item \textbf{Extraterrestrial solar applications only.} The solar--matter interaction quantities in the database are calculated with the extraterrestrial solar spectrum (AM0). They are therefore directly representative of solar irradiation outside the Earth's atmosphere and are particularly suitable for space-based applications involving solar--matter interaction, such as solar cells, photovoltaic systems and space-debris removal systems operating in space. For terrestrial applications, atmospheric absorption and scattering modify the spectral distribution and reduce the irradiance at specific wavelengths, so the solar spectrum at the Earth's surface differs from AM0. The database results should thus be used with low confidence for terrestrial applications. In the future, standard terrestrial spectra such as AM1.5 could be incorporated into the database, so that the same optical properties and calculation framework could be applied to both extraterrestrial and atmosphere-modified solar irradiation, extending its applicability to ground-based solar-energy applications.
\end{itemize}

\section*{Data Availability}
The SpaderOptics database is available on Zenodo at \fld{https://doi.org/10.5281/zenodo.23643910}~\cite{spaderoptics_data} under a CC-BY-4.0 licence. The deposit contains the database file \fld{SpaderOptics.mat} (MATLAB v7.3/HDF5 format, about 1.4~GB on disk), the file \fld{Resampling_error_analysis_results.txt} with the resampling error of every source, and a \fld{README.md} describing the files and the structure of the database. The database is organized by material and by literature source \cite{Hagemann1975-sg,Arakawa1977-pb,Larruquert2013-nx,Amotchkina2020-dh,Aspnes1983-gr,Burnett2016-vx, Barnes1979-hy,Icenogle1976-cy,Jellison1992-et,Li1980-nj,Nunley2016-fr,Ciesielski2018-as,Arakawa1981-im,Brannon1985-wb,Cunningham2011-js,French2009-dv,Kawka2001-zd,Kumar2011,Philipp1986-fq,Sahin2019-ci,Smith1975-og,Zhang1998-hy,Beadie2015-ka,Bodurov2016-ka,Nyakuchena2023-wn,Sultanova2009-bl,Szczurowski2013,Tsuda2018-gg,Zhang2020-vr,Chandler-Horowitz2005-ei,Schinke2013-ud,Edwards1980-in,Green1995-rt,Green2008-mu,Salzberg1957-sl,Schinke2015-py,Shkondin2017-ym,Sik1998-sx,Vuye1993-se,Wang2025-cv,Franta2013-qz,Karaman2025-ua,Pierce1972-rc,Shaffer1971-za,Fischer2017-yw,Wang2013-mn,Singh1971-fc,Larruquert2011-av,Ghosh1999-cs,Radhakrishnan1951-gd,Arosa2020-ny,Franta2016-jo,Malitson1965-sc,Kitamura2007-qf,Gao2013-yj,Kischkat2012-td,Gao2012-cp,Rodriguez-de_Marcos2016-pw,Pfluger1984-wo,Synowicki2004-fq,Wood1982-gz,Babar2015-ir,Choi2020-nm,Ciesielski2017-jj,Jiang2016-zc,Johnson1972-fh,Stahrenberg2001-nk,Werner2009-me,Windt1988-tn,Wu2014-nb,Yang2015-gt,Boidin2016-sz,Franta2015-li,Malitson1962-ub,Zhukovsky2015-cc,Klinavicius2025-jx,Gao2011-bk,Magnozzi2019-bh,Olmon2012-yf,Ordal1987-bs,Rosenblatt2020-pt,Yakubovsky2017-ou,Ordal1985-zb,Mash1973-zb,Palm2018-ed,Bond1965-vl,DeVore1951-tw,Jolivet2023-ds,Sarkar2019-on,Siefke2016-iq,Ciesielski2018-Au}, and its structure is described in the Data Records section. The input refractive indices were taken from the RefractiveIndex.INFO database~\cite{Polyanskiy2024-cv} and the extraterrestrial solar spectrum from the ASTM E490 AM0 reference spectrum~\cite{ASTM_E490}. These data belong to their original authors and are not redistributed in our deposit; they are available from their respective repositories. The database was built from the versions of these data retrieved on 4 May 2026.

\section*{Code Availability}
The code used to build the database, the SpaderOptics App and the scripts and instructions for loading the database with MATLAB and Python are available on GitHub (\fld{https://github.com/V-Alexopoulou/SpaderOpticsCodes}) and are archived on Zenodo at \fld{https://doi.org/10.5281/zenodo.23645807}~\cite{spaderoptics_code} under the GNU General Public License v3.0. The code is organized in seven folders: Steps~1--2 import the refractive indices, Step~3 calculates the solar--matter interaction, Step~4 calculates the emissivity and the equilibrium temperatures, Step~5 assembles the final database, Step~6 contains the SpaderOptics App, and Step~7 contains the Python loading script and the MATLAB and Python README files that describe how to load the database and retrieve data, either through MATLAB or through Python. Steps~1--5 must be run in sequence without clearing the workspace. The code was developed and tested with MATLAB R2025b; the Python loader requires \fld{numpy} and \fld{h5py}. The YAML files of RefractiveIndex.INFO \cite{Polyanskiy2024-cv} and the AM0 solar spectrum \cite{ASTM_E490} are not included in the repository, because they were created by other authors; our repository README files list the exact YAML and AM0 spectrum files to download, where to place them, and the retrieval date (4 May 2026), in order for the user to be able to run Steps~1--7 and reproduce our database and our App. Users who only want to use the database can skip Steps~1--5 and run the App (Step~6) directly with the file from Zenodo.

\section*{Acknowledgements}
The authors gratefully acknowledge Professor Antonios E. Giannakopoulos, School of Applied Mathematical and Physical Sciences, National Technical University of Athens, for his fruitful discussions and guidance.

\section*{Funding}
This publication was funded by the European Union under grant agreement number 101223931. Its contents are the sole responsibility of the SPADER consortium and do not necessarily reflect the views of the European Commission.

\bibliographystyle{sn-nature}
\bibliography{references}

\end{document}